\documentclass[10pt,letterpaper]{article}
\usepackage[top=0.85in,left=2.75in,footskip=0.75in]{geometry}

\usepackage{amsmath,amssymb}

\usepackage{changepage}

\usepackage[utf8x]{inputenc}

\usepackage{textcomp,marvosym}

\usepackage{cite}

\usepackage{nameref,hyperref}
\usepackage{float}
\usepackage[version=4]{mhchem}
\usepackage[right]{lineno}

\usepackage{microtype}
\DisableLigatures[f]{encoding = *, family = * }

\usepackage[table]{xcolor}

\usepackage{array}

\newcolumntype{+}{!{\vrule width 2pt}}

\newlength\savedwidth

\newcommand\thickhline{\noalign{\global\savedwidth\arrayrulewidth\global\arrayrulewidth 2pt}%
\hline
\noalign{\global\arrayrulewidth\savedwidth}}

\raggedright
\usepackage[aboveskip=1pt,labelfont=bf,labelsep=period,justification=raggedright,singlelinecheck=off]{caption}

\makeatletter
\renewcommand{\@biblabel}[1]{\quad#1.}
\makeatother

\date{}

\usepackage{lastpage,fancyhdr,graphicx}
\usepackage{epstopdf}
\fancyheadoffset[L]{2.25in}
\fancyfootoffset[L]{2.25in}

\begin{document}
\vspace*{0.2in}

\begin{flushleft}
{\Large
\textbf\newline{Rare-Event Sampling of Epigenetic Landscapes and Phenotype Transitions} 
}
\newline
\\
Margaret J. Tse\textsuperscript{1},
Brian K. Chu\textsuperscript{1},
Elizabeth L. Read\textsuperscript{1,2*}
\\
\bigskip
\textbf{1} Dept. of Chemical Engineering \& Materials Science, University of California, Irvine, California, USA
\\
\textbf{2} Center for Complex Biological Systems, University of California, Irvine, California, USA
\\
\bigskip

* elread@uci.edu

\end{flushleft}
\section*{Abstract}

Stochastic simulation has been a powerful tool for studying the dynamics of gene regulatory networks, particularly in terms of understanding how cell-phenotype stability and fate-transitions are impacted by noisy gene expression. However, gene networks often have dynamics characterized by multiple attractors. Stochastic simulation is often inefficient for such systems, because most of the simulation time is spent waiting for rare, barrier-crossing events to occur. We present a rare-event simulation-based method for computing epigenetic landscapes and phenotype-transitions in metastable gene networks. Our computational pipeline was inspired by studies of metastability and barrier-crossing in protein folding, and provides an automated means of computing and visualizing essential stationary and dynamic information that is generally inaccessible to conventional simulation. Applied to a network model of pluripotency in Embryonic Stem Cells, our simulations revealed rare phenotypes and approximately Markovian transitions among phenotype-states, occurring with a broad range of timescales. The relative probabilities of phenotypes and the transition paths linking pluripotency and differentiation are sensitive to global kinetic parameters governing transcription factor-DNA binding kinetics. Our approach significantly expands the capability of stochastic simulation to investigate gene regulatory network dynamics, which may help guide rational cell reprogramming strategies. Our approach is also generalizable to other types of molecular networks and stochastic dynamics frameworks.

\section*{Author summary}

Cell phenotypes are controlled by complex interactions between genes, proteins, and other molecules within a cell, along with signals from the cell's environment. Gene regulatory networks (GRNs) describe these interactions mathematically. In principle, a GRN model can produce a map of possible cell phenotypes and phenotype-transitions, potentially informing experimental strategies for controlling cell phenotypes. Such a map could have a profound impact on many medical fields, ranging from stem cell therapies to wound healing. However, analytical solution of GRN models is virtually impossible, except for the smallest networks. Instead, time course trajectories of GRN dynamics can be simulated using specialized algorithms. However, these methods suffer from the difficulty of studying rare events, such as the spontaneous transitions between cell phenotypes that can occur in Embryonic Stem Cells or cancer cells. In this paper, we present a method to expand current stochastic simulation algorithms for the sampling of rare phenotypes and phenotype-transitions. The output of the computational pipeline is a simplified network of a few stable phenotypes, linked by potential transitions with quantified probabilities. This simplified network gives an intuitive representation of cell phenotype-transition dynamics, which could be useful for understanding how molecular processes impact cellular responses and aid interpretation of experimental data.


\section*{Introduction}

In multicellular organisms, differentiation of pluripotent stem cells into tissue-specific cells was traditionally considered to be an irreversible process. The discovery of cell reprogramming revealed that a the identity of a cell is not irreversibly stable, but rather plastic and amenable to control by perturbation of gene regulatory interactions—for example, through over-expression of key transcription factors~\cite{takahashi_induction_2007}. Cellular plasticity has also been observed in other contexts, where cells appear to spontaneously transition among phenotypically distinct states. For example, in embryonic stem cells, expression levels of key transcription factors show dynamic heterogeneity, which is thought to enable diversification of the population prior to lineage commitment~\cite{abranches_stochastic_2014,dietrich_stochastic_2007,kalmar_regulated_2009,singh_heterogeneous_2007,ohnishi_cell--cell_2014}. This heterogeneity may result at least in part from stochastic state-transitions between functionally distinct, metastable subpopulations~\cite{kalmar_regulated_2009,singer_dynamic_2014,filipczyk_network_2015,hormoz_inferring_2016}. Stochastic state-transitions have also been proposed to play a role in cancer, by enabling cancer stem cells to arise \emph{de novo} from non-stem subpopulations~\cite{gupta_stochastic_2011}, or by enabling cells to reversibly transition to a drug-tolerant phenotype~\cite{sharma_chromatin-mediated_2010}. In microbial systems, stochastic phenotype switching has been identified as a survival mechanism for populations subjected to fluctuating environments~\cite{acar_stochastic_2008,balaban_bacterial_2004}.

Mathematical modeling has provided a basis for understanding how gene regulatory mechanisms and network interactions control cellular identity, stability, and phenotype-transitions. These approaches yield a quantitative means of reinterpreting the long-standing conceptual framework known as Waddington's epigenetic landscape~\cite{waddington_strategy_1957,bhattacharya_deterministic_2011,wang_quantifying_2011,huang_molecular_2012}. In a mathematical framework, the ``valleys" in the landscape that stabilize cell identities within distinct lineages correspond to attractor basins of a high-dimensional nonlinear dynamical system~\cite{huang_cell_2005}. The nonlinearity results from positive feedback in transcriptional regulation and epigenetic barriers to chromatin remodeling, for example. These feedback mechanisms give rise to multiple, stable (or metastable) phenotype-states accessible to a given genome. Given the ``bursty'' nature of gene expression and ever-present molecular fluctuations in the cell~\cite{elowitz_stochastic_2002,kaern_stochasticity_2005}, an active area of research is in modeling the effects of so-called intrinsic noise on gene regulatory network (GRN) dynamics. These mathematical models support the idea that intrinsic noise can drive stochastic phenotype-transitions~\cite{aurell_epigenetics_2002,sasai_stochastic_2003,feng_2012,tse_dna-binding_2015,ge_stochastic_2015}, which, though likely to be exceedingly rare in general cellular contexts, may explain the heterogeneity observed in embryonic stem cells where epigenetic barriers appear to be lowered~\cite{chang_transcriptome-wide_2008}.

Mathematical models of GRN dynamics that treat stochastic molecular processes are often formulated as probabilistic Master Equations, in which the system evolves probabilistically over a discrete state-space of molecular species and configurations according to a defined set of biochemical reaction rules.  Another common framework is that of a coupled system of ODEs describing the expression levels of genes in the network, with the inclusion of additive noise terms. The Master Equation framework is well-suited to studying how ``local" stochastic molecular events (e.g., transcription factors interacting with DNA or chromatin state-transitions near promoters) impact ``global'' dynamics of phenotype stability and state-switching~\cite{sasai_time_2013,feng_2012,zhang_stem_2014,tse_dna-binding_2015,ge_stochastic_2015}. These molecular fluctuations affecting promoter activity have been shown to significantly impact the structure of epigenetic landscapes, motivating the use of Master Equation-based approaches. That is, the number and stability of phenotype-states accessible to a given GRN varies depending on the kinetic parameters governing these fluctuations~\cite{feng_2012,tse_dna-binding_2015,chu_markov_2017}. Furthermore, ODE or ``mean-field" models that average over these fluctuations can show qualitatively different landscape features~\cite{lipshtat_genetic_2006,schultz_extinction_2008,ma_small-number_2012}. 

Master Equation approaches face the well-known challenge of the ``Curse-of-Dimensionality'', as solving them requires enumeration of a state-space that grows exponentially with the number of molecular species in the network. For this reason, discrete stochastic models of GRNs are often studied by stochastic Monte Carlo simulation, via the Gillespie algorithm~\cite{gillespie_exact_1977}. However, stochastic simulation can also be problematic: in systems with metastability, such as GRNs, stochastic simulation becomes highly inefficient. Transitions between metastable states are rare events (i.e., rare relative to the timescale of fluctuations within a metastable attractor basin), and thus difficult or impossible to observe. Often, these rare events are precisely the events of interest, such as in GRNs where infrequent state-transitions represent critical cell-fate transitions. 

Rare-event sampling algorithms are designed to overcome these challenges, by redirecting computational resources towards events of interest, while maintaining statistical accuracy to global system dynamics~\cite{allen_forward_2009,zuckerman_weighted_2017}. In this work, we present a rare-event simulation-based method for computing and analyzing epigenetic landscapes of stochastic GRN models. We combine rare-event methods with coarse-graining and analysis by Transition Path Theory--adopted from the field of Molecular Dynamics of protein folding~\cite{noe_constructing_2009}--and show that this unified framework provides an automated approach to map epigenetic landscapes and transition dynamics in complex GRNs. The method quantifies the number of metastable phenotype-states accessible to a GRN, calculates the rates of transitioning among phenotypes, and computes the likely paths by which transitions among phenotypes occur. We apply the method to a model of pluripotency in mouse Embryonic Stem Cells. Our results reveal rare sub-populations and transitions in the network, demonstrate how global landscape structure depends on kinetic parameters, and reveal irreversibility in paths of differentiation and reprogramming. Our approach is not limited to gene regulatory networks; it is generalizable to other stochastic dynamics frameworks and is thus a potentially powerful tool for computing global dynamic landscapes in areas such as signal-transduction, population dynamics, and evolutionary dynamics.


\section*{Methods}

\begin{figure}[H]\includegraphics[width=\textwidth]{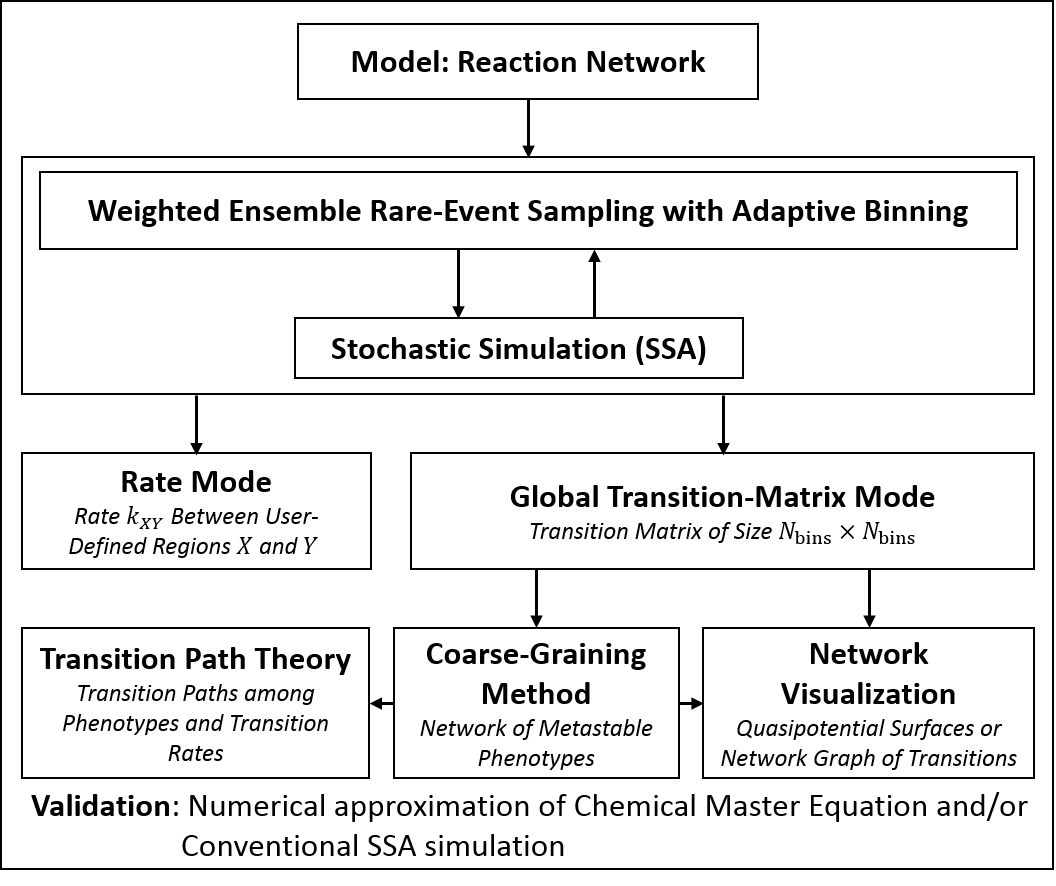}
\caption{{\bf Computational Pipeline for Rare-Event Sampling of Epigenetic Landscapes and Phenotype Transitions.} The input to the computational pipeline is a reaction network model of gene regulatory network dynamics. Stochastic simulations are performed using SSA~\cite{gillespie_exact_1977} and Weighted Ensemble rare-event sampling~\cite{huber_weighted-ensemble_1996}. The WE method can be run in two modes: \emph{Rate Mode} computes the rate of transitioning between two user-defined regions of interest with high accuracy. \emph{Transition-Matrix Mode} computes the pairwise transition probabilities among $N_\mathrm{bins}$ adaptively defined sampling bins that span the system state-space. Further visualization and analysis of the transition-matrix can be performed, including automatic designation of metastable phenotypes via the coarse-graining framework~\cite{prinz_markov_2011} and identification of likely transition paths~\cite{noe_constructing_2009}.
}
\label{fig:Graphical_pipeline}
\end{figure}

A graphical overview of the computational pipeline presented in this paper can be found in Fig \ref{fig:Graphical_pipeline}.

\subsection*{Gene Regulatory Network Models}
We demonstrate the rare-event sampling method for two representative GRN models. A small, two-gene network serves as a model system to validate the simulations. We then apply the method to a more complex model of pluripotency in mouse Embryonic Stem Cells (mESCs). 
\subsubsection*{Exclusive Mutual Inhibition, Self-Activation Model}
The Exclusive Mutual Inhibition, Self-Activation (ExMISA) model is a two-gene network representing an archetypal motif at cell-fate branch points~\cite{huang_reprogramming_2009,graf_forcing_2009}. Each gene, denoted generically as $A$ or $B$, encodes a transcription factor that activates its own transcription and represses transcription of the other gene.
We adopt previous conventions~\cite{kepler_stochasticity_2001,sasai_stochastic_2003,feng_2012} for stochastic GRN dynamic models. The full list of biochemical reactions and parameters can be found in the Supplement, \nameref{S1_File} and \nameref{S:ExMISAparam}. The model encompasses stochastic birth/death processes for transcription factor production and degradation, and stochastic binding and unbinding of transcription factors to DNA regulatory/promoter regions; the binding-states of these regions governs the production rate. Each transcription factor is assumed to bind to DNA as a homodimer, giving cooperative regulation. In the ``exclusive'' network variant, transcription factors compete for binding sites on DNA (only one transcription factor dimer can be bound to a gene's promoter at a time). The discrete state-vector, which completely describes the state of the system, is given by $\mathbf{x}=[A_{ij},B_{ij},n_a, n_b]$. $A_{ij}$ and $B_{ij}$ represent the three possible promoter binding-states for each gene (i.e., $A/B_{00},A/B_{10},A/B_{01}$ denote unbound, activator-bound, or repressor-bound states). The copy-numbers of expressed protein transcription factors are denoted by $n_a$ and $n_b$ for products of gene $A$ and $B$, respectively, and may in principle take any nonnegative integer value. All processes related to transcription, translation, and assembly are subsumed into a single protein birth reaction. For genes in state $A/B_{ij}$, this production occurs with rate constant $g_{ij}$. The production rate is high when the promoter is bound only by the activator (its own product). Otherwise, if unbound or repressor-bound, a low ``basal'' rate of expression is assumed, i.e. $g_{00}=g_{01}<g_{10}$. Degradation of protein products occurs with rate $k$, and stochastic binding/unbinding of transcription factors to DNA occur with $h$ and $f$, respectively. The model is symmetric, with equivalent parameters for the two genes. 

\subsubsection*{Pluripotency Network Model}
The pluripotency network model of mESCs was developed by Zhang and Wolynes~\cite{zhang_stem_2014} on the basis of experimental literature and previous models. The 8-gene network shares the same stochastic reaction framework as the ExMISA model. The genes (NANOG, OCT4, SOX2, GCNF, KLF4, PBX1, GATA6, and CDX2) suppress and activate each other through homo- and heterodimers of their encoded transcription factors (OCT4 and SOX2 form a heterodimer; all other regulatory interactions occur via homodimers). Binding of transcription factors to promoters is not exclusive. The model has five kinetic parameters: $g_{\mathrm{on}}$, $g_{\mathrm{off}}$, $h$, $f$, and $k$, corresponding to the rate of gene expression in the activated state, the rate of gene expression in the un-activated state, binding of transcription factors to DNA, unbinding of transcription factors from DNA, and transcription factor degradation (or exit from the nucleus). Genes are expressed at the basal rate $g_{\mathrm{off}}$ except when bound by at least one activator and no repressor, in which case they are expressed with rate $g_{\mathrm{on}}$. The exception to this logic rule is NANOG, which must be bound by the the KLF4 and PBX1 transcription factor homodimers and the heterodimer OCT4-SOX2 to be activated. Overall, these interactions lead to a total of 396 biochemical reactions, with a total of 88 ``species" (counting 80 distinct gene promoter configurations and 8 protein species).  The complete logic rules and list of reaction rate parameters can be found in the Supplement (\nameref{S1_File}, \nameref{S:pluripotency_network}, and \nameref{S:pluripotency_network_param}).

\subsection*{Theoretical Background: the Chemical Master Equation and Stochastic Transition-Matrix}
The mathematical framework of the network models is the discrete Chemical Master Equation (CME)~\cite{gillespie_exact_1977}, which gives the time-evolution of the probability to observe the system in a given state. In vector-matrix form, the CME can be written
\begin{eqnarray}
\label{CME}
\frac{d\mathbf{p}(\mathbf{x},t)}{dt}=\mathbf{K}\mathbf{p}(\mathbf{x},t)
\label{eq:CME}
\end{eqnarray}
where $\mathbf{p}(\mathbf{x},t)$ is the probability over the system state-space ($\mathbf{x}$) at time $t$, and $\mathbf{K}$ is the reaction rate-matrix containing stochastic reaction propensities (diagonal elements $k_{jj}=-\sum_i k_{ij}$, i.e., columns sum to 0). Equation \ref{CME} assumes a well-mixed system of reacting species, and assumes that the technically infinite state-space described by $\mathbf{x}$ (containing molecular species numbers/configurations) may be limited to some finite number of ``reachable" states, (i.e., with non-negligible probability) for an enumeration of $N$ states of the system, $\mathbf{K}\in \mathbb{R}^{N\times N}$. The steady-state probability $\mathbf{\pi}(\mathbf{x})\equiv\mathbf{p}(\mathbf{x},t\rightarrow \infty)$ over $N$ states satisfies 
\begin{eqnarray}
\label{eq:KSS}
\mathbf{K}\mathbf{\pi}(\mathbf{x})=\mathbf{0}.
\end{eqnarray}
Thus, $\mathbf{\pi}(\mathbf{x})$ can be obtained from $\mathbf{K}$ as the normalized right-eigenvector corresponding to the zero-eigenvalue.

It is sometimes desirable to work with the time-dependent stochastic transition-matrix $\mathbf{T}(\tau)$ rather than the time-independent stochastic rate matrix $\mathbf{K}$~\cite{prinz_markov_2011}. For example, $\mathbf{T}(\tau)$ may be more amenable to estimation by sampling (as we demonstrate in this work for the pluripotency network, for which $\mathbf{K}$ is impractical to enumerate). For a CME with rate matrix $\mathbf{K}$, $\mathbf{T}(\tau)$ is given by 
\begin{eqnarray}
\label{eq:TMat}
\mathbf{T}(\tau)=\mathrm{exp}(\tau\mathbf{K}^{\mathrm{T}})
\end{eqnarray}
where $\mathrm{exp}$ denotes the matrix exponential. $\mathbf{T}(\tau)\in \mathbb{R}_{0\leq x \leq1}^{N\times N}$ then gives the conditional probability for the system to transition between each pair of states within a lagtime $\tau$. That is, the elements $t_{ij}$ give the probability that the system, if found in state $i$, will then be found in state $j$ at a time $\tau$ later, and rows sum to 1.  Using $\mathbf{T}(\tau)$, the evolution of probability over discrete intervals of the lagtime $\tau$ is given by the Chapman-Kolmogorov equation: 
\begin{eqnarray}
\label{eq:Chap}
\mathbf{p}^T(\mathbf{x},t+k\tau)=\mathbf{p}^\mathrm{T}(\mathbf{x},t)\mathbf{T}^k(\tau).
\end{eqnarray}
Eigenvectors corresponding to dominant eigenvalues of the stochastic transition-matrix are associated with slow system processes. By Perron-Frobenius, for an irreducible stochastic matrix $\mathbf{T}(\tau)$ with eigenvalues $\lambda_i$, there exists $\lambda_1=1$, and all other eigenvalues satisfy $\lvert{\lambda_i}\rvert < 1$. Analogous to Equation~(\ref{eq:KSS}) for $\mathbf{K}$, the steady-state probability can be obtained directly from $\mathbf{T}(\tau)$ according to $\mathbf{\pi}^T(\mathbf{x})=\mathbf{\pi}^T(\mathbf{x})\mathbf{T}(\tau)$, i.e., as the normalized left-eigenvector corresponding to $\lambda_1$. Eigenvalues $\lambda_i$ are related to global system timescales $t_i$ by 
\begin{eqnarray}
\label{eq:timescales}
t_i=-\frac{\tau}{\mathrm{ln} \lvert \lambda_i(\tau) \rvert},
\end{eqnarray}
(with $t_1$ giving the infinite-time, stationary result)~\cite{prinz_markov_2011}. Additionally, the Mean First Passage Time ($\mathrm{MFPT}_{X,Y}$) where $X$ and $Y$ are individual states can be computed using the matrix elements $T_{i,j}$ by~\cite{scherer_pyemma_2015, hoel_introduction_1986}:
\begin{eqnarray}
\label{eq:MFPT}
\mathrm{MFPT}_{X,Y} = \tau \times \Bigg{\{}\begin{aligned} & 0 & X = Y\\
&1 + \sum_z T_{X,Z} \mathrm{MFPT}_{Z,Y} & X\ne Y \end{aligned}.
\end{eqnarray} 

\subsection*{Weighted Ensemble Stochastic Simulation}
Stochastic reaction kinetics can be simulated by the Stochastic Simulation Algorithm (SSA) ~\cite{gillespie_exact_1977}, which produces numerically exact realizations of the CME (Eq ~\ref{eq:CME}). Simulation circumvents the need for enumerating the exceedingly large system state-spaces typical of gene network models, but suffers from inefficiency due to rare events. The Weighted Ensemble (WE) rare-event sampling algorithm~\cite{huber_weighted-ensemble_1996} redistributes computational resources from high-probability regions of state-space to low-probability regions, which tend to be under-sampled in conventional simulation. The method thereby reduces computational effort in sampling rare transitions and improves accuracy of estimating probability density in, e.g., barrier-regions or tails of distributions. The method can be applied to any stochastic dynamics framework; in recent years, it has been widely applied to atom-scale Molecular Dynamics. Details of the methodology are discussed in a recent review~\cite{zuckerman_weighted_2017} and references therein. 

Briefly, the algorithm works as follows: state-space is divided up into bins that span transitions of interest. The number of bins, $N_\mathrm{bins}$, is typically $\mathcal{O}(100)$, and a variety of binning procedures can be used (we use an adaptive procedure described below). Initially, a single simulation trajectory, or ``replica", is assigned a weight of 1 and allowed to freely move within and between bins for a user-defined lagtime $\tau_{\mathrm{WE}}$. After each iteration of $\tau_\mathrm{{WE}}$, a splitting and culling procedure divides and/or combines replicas and their associated weights in such a way as to reach and maintain an equal target number of weighted replicas, $M_{\mathrm{targ}}$, in each bin. Over the course of the simulation, the combined weights of the replicas in a bin (averaged over successive iterations) will evolve toward the probability of the system to reside in that bin. By maintaining the same number of replicas in each bin ($M_\mathrm{targ}$), with weights proportional to probability, the algorithm devotes comparable computational time to low- and high-probability regions. Effectively, the algorithm computes long-time processes on the basis of many short-time simulated trajectories.

\subsubsection*{Adaptive Binning Procedure}
As with other enhanced sampling methods, the WE algorithm requires dividing of state-space into defined sampling regions or ``bins". For high-dimensional systems, discretization poses a challenge because, for an $N$-dimensional, evenly spaced grid, the number of required sampling bins increases exponentially with the number of degrees of freedom. To address this challenge, a variety of Voronoi-polyhedra-based procedures have been developed~\cite{dickson_nonequilibrium_2009,dickson_wexplore:_2014,zhang_weighted_2010}. These methods balance the need to focus simulation toward regions with non-negligible probability, while still enabling capture of rare transitions of interest. In addition to efficiently discretizing high-dimensional spaces, the methods have the benefit of requiring little to no \emph{a priori} knowledge of system dynamics (e.g., of the locations of regions of interest, or of appropriate progress coordinates for transitions). We utilize an adaptive binning procedure from ref.~\cite{zhang_weighted_2010}. Each bin (of user-defined number $N_\mathrm{bins}$) is a Voronoi polyhedron with a generating node; the bin is defined as the region of state-space encompassing all points closer to the generating node than to nodes of any other region. After each lagtime $\tau_{\mathrm{WE}}$, new Voronoi regions are generated by successively selecting $N_\mathrm{bins}$ node-positions from the current replica positions in a way that maximizes the Euclidean distance between them. By this procedure, over the course of the simulation, bins spread to encompass all areas of state-space reached by any simulated trajectory. After sufficient iterations, the bin positions stop spreading to new areas but continue to fluctuate. The procedure is shown by representative simulations in \nameref{S:Voronoi_movement}. 

\subsubsection*{Computation of Transition Rates}
One important output of WE sampling is the quantitative rate of transitions between regions of interest, which may be difficult or impossible to estimate from conventional simulation. WE sampling may be run in different modes, depending on whether the sought-after information concerns a specific transition of interest, or a more global picture of system dynamics, i.e., encompassing approximate rates of transitions among many system states. We term the two modes ``rate" mode and ``global transition-matrix'' mode. The former can deliver a more accurate estimate for a particular state-transition, while the latter can yield a more comprehensive, but approximate, measure of global system dynamics.

In rate mode, the user specifies two regions of interest, $X$ and $Y$, The flux of probability into/out of regions of interest can be estimated by recording the amount of weight transferred at the end of each simulation iteration. The mean first passage time of transitions from $X$ to $Y$ ($\mathrm{MFPT}_{X,Y}$) is given in general by the inverse of probability flux from $X$ to $Y$. In practice, we apply a``labeling" scheme~\cite{suarez_simultaneous_2014,dickson_separating_2009}, where each replica is labeled as belonging to either set $\mathcal{S}_X$ or $\mathcal{S}_Y$ according to its history, i.e., whether it most recently visited region $X$ or $Y$, respectively. The summed weight of all replicas in $\mathcal{S}_X$ is given by $P_{\mathcal{S}_X}$, and $P_{\mathcal{S}_X}+P_{\mathcal{S}_Y}=1$ satisfies probability conservation. Then, 
\begin{eqnarray}
\label{eq:MFPT_WE}
\mathrm{MFPT}_{X,Y}=\frac{\overline{P}_{\mathcal{S}_X}^{\mathrm{SS}}}{\overline{\Phi}^{\mathrm{SS}}(Y\vert\mathcal{S}_X)}
\end{eqnarray}
where $\overline{\Phi}^{\mathrm{SS}}(Y\vert \mathcal{S}_X)$ is the average probability flux from $\mathcal{S}_X$ into $Y$ at steady-state, which is measured by the weight of $\mathcal{S}_X$-labeled replicas entering $Y$ during the simulation after convergence to steady-state. The labeling scheme enables accurate estimates, including for non-Markovian transitions. For Markovian transitions well-described by a single rate-constant, $k_{X,Y}=1/\mathrm{MFPT}_{X,Y}$. 

\subsubsection*{Computation of Network Transition-Matrix}
Running WE in transition-matrix mode enables visualization and analysis of global system dynamics on the basis of a single simulation, and requires no designation of regions of interest. In this mode, the previously-converged Voronoi bins are fixed, and simulations are used to estimate a coarse-grained stochastic transition-matrix $\widetilde{\mathbf{T}}(\tau)$ of size $N_\mathrm{bins}\times N_\mathrm{bins}$. The coarse-grained $\widetilde{\mathbf{T}}(\tau)$ approximates the true dynamics over the full state-space, as given by $\mathbf{T}(\tau)$. Thus, the procedure enables estimation of the global transition-matrix (and subsequent analysis) in systems where enumeration of states is not feasible. 
To estimate $\widetilde{\mathbf{T}}(\tau)$, the weight transferred between bins is recorded at each iteration, and the elements of the transition-matrix are estimated according to~\cite{suarez_simultaneous_2014}:
\begin{eqnarray}
\label{eq:counts}
\widetilde{T}_{i,j}=\frac{\langle w_{i,j}\rangle_2}{\langle w_{i}\rangle}
\end{eqnarray}
where $\langle w_{i,j}\rangle_2$ is the average weight transferred from bin $i$ to bin $j$ over the iteration time $\tau_{WE}$ (counting only after at least 2 transitions, and averaging over multiple iterations) and $\langle w_i \rangle$ is the average population (summed weight) in bin $i$. By construction, this is a row-stochastic transition-matrix with state-space ``resolution'' determined by $N_\mathrm{bins}$ (each state in the full state-space sampled by the simulation is assigned to its nearest neighboring Voronoi node). The lagtime $\tau$ of the transition-matrix corresponds to the sampled WE-time $\tau_{\mathrm{WE}}$.  However, use of $\widetilde{\mathbf{T}}(\tau)$ to compute system dynamics imposes a Markovian approximation, by which equilibration of replicas within bins is assumed to be rapid on the timescale of $\tau$, and hops between states (i.e. bins) are memoryless. As such, while this mode of simulation has the advantage of acquiring a holistic view of global system dynamics, it has the disadvantage of introducing a Markovian approximation.

\subsection*{Coarse-Graining Procedure to Classify Phenotype-States}
While the sampled $N_\mathrm{bins}\times N_\mathrm{bins}$ transition-matrix provides a global approximation of the epigenetic landscape and state-transitions, we apply a method to further coarse-grain dynamics, known as the Markov State Model framework~\cite{prinz_markov_2011,noe_constructing_2009,chu_markov_2017}. This automated procedure produces a highly simplified representation of global dynamics in terms of a few (generally $<10$) clustered sets and the transitions among them. Such highly-reduced models can be beneficial in terms of human intuition of system dynamics, comparison to experiments, and--in this application--automated designation of dynamic phenotype-states. The method utilizes the concept of metastability, i.e., system states that experience relatively fast transitions among them are clustered together into the same coarse-grained set.  Collectively, the coarse sets experience relatively rare inter-cluster transitions and frequent intra-cluster transitions. We employ the metastability concept as a definition of cell phenotype, reasoning that a phenotype should be a relatively stable attribute of a cell, and stochastic inter-phenotype transitions should be relatively rare. In practice, we employ the Markov State Model framework to further reduce the sampled row-stochastic transition-matrix $\widetilde{\mathbf{T}}(\tau)$ from size $N_\mathrm{bins}\times N_\mathrm{bins}$ down to $C\times C$, where $C$ is the number of coarse-grained clusters. As the Markov State Model (MSM) is itself a stochastic transition-matrix on a coarse-grained space, it implies a more severe Markovian approximation. It provides a way to describe global system dynamics in a highly simplified way while maintaining high accuracy to the slowest system dynamics as sampled by $\widetilde{\mathbf{T}}(\tau)$. In previous work, we demonstrated the application of this coarse-graining approach to automatically designate phenotypes in small gene networks~\cite{chu_markov_2017}; here, we extend the applicability of the coarse-graining to large, complex networks by combining it with rare-event sampling.

The coarse-graining procedure is a spectral clustering method based on the Perron Cluster Cluster Analysis (PCCA+) algorithm~\cite{roblitz_fuzzy_2013}, which optimizes the (nearly)-block-diagonal structure of $\widetilde{\mathbf{T}}(\tau)$ for systems with metastability. The signature of such metastability is a separation-of-timescales for intra- and inter-basin dynamics, which may be seen as gaps in the eigenvalue spectrum~\cite{prinz_markov_2011}. As noted above, $\mathbf{T}(\tau)$ (or its sampled counterpart, $\widetilde{\mathbf{T}}(\tau)$) has $\lambda_1=1$, corresponding to the infinite time-limit. If a set of $m$ dominant eigenvalues exists, such that for decreasing eigenvalues $\lambda_{i}\lessapprox 1$, $i\in \{2,...,m\}$, and a gap is present, $\lambda_j <<\lambda_m$ for $j>m$, this indicates the presence of $m$ slow-timescale processes in the system, and further indicates that $\widetilde{\mathbf{T}}(\tau)$ may be re-ordered to give $m$ nearly-uncoupled blocks. In practice, the algorithm attempts to find a coarse-graining onto $C$ clusters, where $C$ may be user-defined, or may be determined algorithmically, e.g., according to the spectral gap~\cite{roblitz_fuzzy_2013}. Here, we choose $C$ clusters, where the last significant gap in the spectrum is seen between $\lambda_C$ and $\lambda_{C+1}$. For the GRNs studied here, this corresponds to choosing $C$ such that $\lambda_C/\lambda_{C+1}>10$. 

\subsubsection*{Transition Path Analysis}
The coarse-grained model of system dynamics given by the MSM enables estimation of the ensemble of dominant transition paths among phenotypes, along with their relative probabilities. We adopt methods from Transition Path Theory according to Noe, \emph{et al}.~\cite{noe_constructing_2009} (details therein). Briefly, $\widetilde{\mathbf{T}}(\tau)$ can be used to compute the effective flux of trajectories, along any edge in the coarse-grained network, contributing to transitions between states $X$ and $Y$ (where these designated states correspond to one or more coarse-grained phenotype-states produced by the MSM). A pathway decomposition algorithm on the matrix of effective fluxes for $X \rightarrow Y$ transitions then yields a set of dominant pathways and the relative contribution of each to the overall flux. Each state in the MSM is analogous to a cell phenotype, and transition path analysis is used to identify parallel phenotype transition paths and the relative rates of transitioning between phenotypes.

\subsection*{Visualization of Epigenetic Landscapes}
Both the sampled transition-matrix $\widetilde{\mathbf{T}}(\tau)$ and the coarse-grained MSM encode stationary and dynamic information about global dynamics--that is, they quantify the epigenetic landscape. For visualization, we use Gephi graph visualization software \cite{bastian_gephi:_2009} using the Force Atlas algorithm. Every circle (or node) in the graph corresponds to a sampling bin or to a coarse-grained phenotype, and the area of a circle is proportional to its relative steady state probability according to $\ln (\gamma P_{SS})$, where $P_{SS}$ is the steady state probability of the node and $\gamma$ is a constant chosen to improve visibility of low probability regions of the landscape. Lines between circles (edges) correspond to transitions between sampling regions or coarse-grained phenotype. Their thickness and coloring correspond to their relative transition probability and source state, respectively. 

\subsection*{Validation: Numerical Solution of the Chemical Master Equation}
To validate the simulation method, we compare the simulated dynamics to the numerical solution to the CME. We choose the parameters of the ExMISA model in such a way as to restrict the effective state-space, so that a numerical solution of the CME is tractable. Building the reaction rate matrix $\mathbf{K}\in \mathbb{R}^{N\times N}$ requires enumeration of $N$ system states. In general, if a system of $S$ molecular species has a maximum copy number per species of $n_{max}$, then  $N\approx n_{max}^S$. In the ExMISA model, the state-vector is given by $\mathbf{x}=[A_{ij},B_{ij},n_a, n_b]$. For enumeration, we neglect states with protein copy-numbers larger than a cutoff value which exceeds $g_{10}/k$ (corresponding to the average number of transcription factors maintained in the system from a gene while in its active state). For example, with model parameters $g_{10}=18$ and $k=1$, we truncate at $n_{a,max}=n_{b,max}=41$ and assume that probability flux between states with $n_a$, $n_b \le 41$ and states with $n_a$, $n_b>41$ is assumed to be 0 (i.e., the boundaries of the state-space are reflective). Including the gene-binding states, this gives $N = 3 \times 3 \times 42 \times 42 = 15876$ states. This size is tractable for complete solution of the CME using matrix methods in MATLAB~\cite{noauthor_matlab_nodate}. {\color{black}This truncation of the state-space introduces a small approximation error (see \nameref{s:FSP_error})}. 

The pluripotency network has 8 genes with copy numbers of $\mathcal{O}(10^3)$ (determined by the parameters $g_{on}/k=3900$). The number of distinct binding-promoter states for each gene are $16,32,8,8,2,8,4,\text{ and }2$ for GATA6, NANOG, CDX2, OCT4, SOX2, KLF4, GCNF, PBX1, respectively {(see \color{black} \nameref{S:pluripotency_network})}. Together these combinations enumerate a state-space of $N>10^{30} \approx 1000^8 \times 16 \times 32 \times 8 \times 8 \times 2 \times 8 \times 4 \times 2$. This size precludes solution of the CME, and we instead estimate the dynamics by WE sampling. Where possible, we validate the WE-sampling results by ``conventional'', i.e., by direct simulation using SSA.

\subsubsection*{Validation of Coarse-Grained Models}
 To check the validity of the coarse-grained MSM as a representation of the global dynamics, we use the Chapman-Kolmogorov test to compare the relaxation curves of the coarse-grained system to those found through direct SSA following Equation \ref{eq:Chap}~\cite{prinz_markov_2011}. If the coarse-graining is appropriate, the relaxation curves of the MSM probabilities will match the relaxation profile of long conventional (direct SSA) simulations initiated within each coarse-grained phenotype. Transition paths through the coarse-grained phenotype network are validated, where possible, against conventional SSA simulation.

\subsection*{Implementation and Software}
Stochastic Gillespie (SSA) simulations were carried out using BioNetGen~\cite{faeder_rule-based_2009}. WE sampling was implemented with in-house software code written in MATLAB. Simulations were run on the high performance computing cluster (HPC) at the University of California, Irvine, and parallelization of BioNetGen SSA simulations was performed using the Sun Grid Engine scheduler. The coarse-graining procedure and transition path analysis was implemented in python scripts, adapted from MSMBuilder~\cite{harrigan_msmbuilder:_2017} and Pyemma~\cite{scherer_pyemma_2015}, respectively. Transition-matrix and MSM visualization was carried out using Gephi software and the Force Atlas layout\cite{bastian_gephi:_2009}. All simulation parameters can be found in the supplement \nameref{S:WE_param}. 

\section*{Results}
\subsection*{Rare States and Transitions in Gene Regulatory Networks are Accessible by Rare-Event Sampling}
\begin{figure}[H]\includegraphics[width=\textwidth]{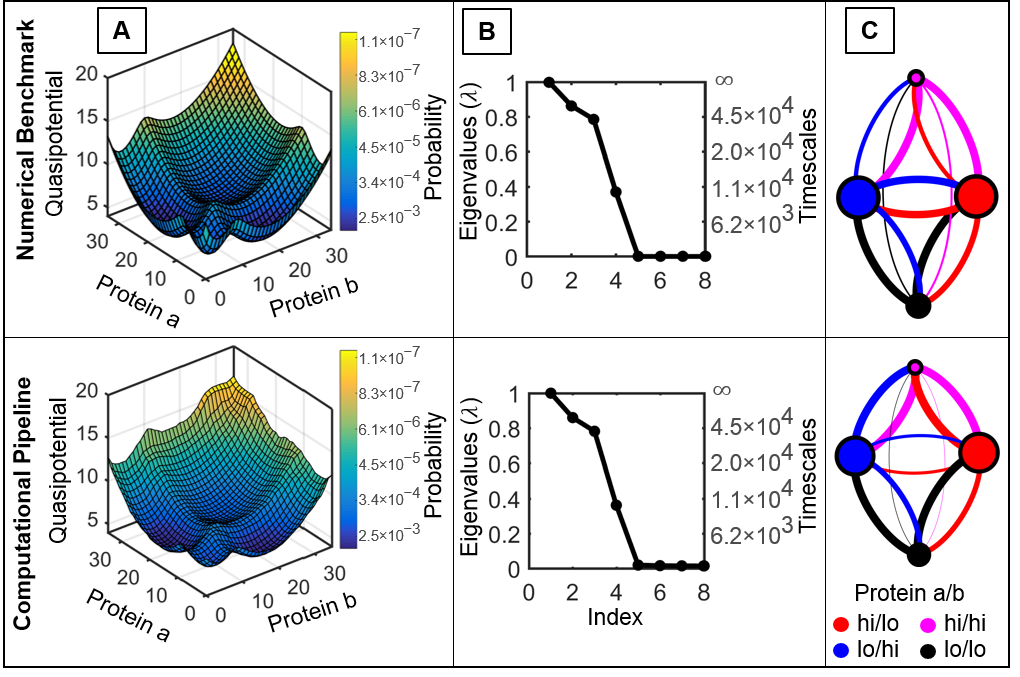}
\caption{{\bf Simulation Results Show Good Agreement with a Theoretical Benchmark for the 2-gene ExMISA (Mutual Inhibition, Self-Activation) Cell-Decision Circuit.} The Chemical Master Equation for the 2-gene model, ExMISA, was solved numerically (see Methods) (top) and compared to simulation results from the computational pipeline presented in this paper (bottom). Shown for each are the Quasipotential Landscape (A), Eigenvalue Spectrum (B), and Markov State Model (C). (A) Quasipotential landscapes of the ExMISA network projected onto the two protein coordinates. Deep blue regions denote low potential (high probability) and yellow denote high potential (low probability). The four visible basins in both correspond to combinations of lo/hi expression for the two genes $A$ and $B$. (For both rows, quasipotential surfaces estimated over discrete states/bins are smoothed for visualization). WE sampling captured both the basin structure and low probability edge and barrier regions. (B) Eigenvalue spectra and corresponding computed global transition timescales. Gaps in the eigenvalue spectrum indicate separation of timescales, i.e., the presence of metastability. C) Four-phenotype coarse-grained models automatically generated from the clustering algorithm (see Methods). Each colored circle represents a cell phenotype, sized proportionally to its probability. Edges are inter-phenotype transitions (colored by source-state, with width proportional to probability). The full CME and simulation pipeline identify similar metastable phenotype networks (see \nameref{S:CG_partitioning} for details).}
\label{fig:MISAEx}
\end{figure}

We first apply the computational pipeline to a small two-gene model (the exclusive Mutual Inhibition, Self-Activation model, ExMISA, see Methods), exhibiting an archetypal motif for cell fate-decisions\cite{graf_forcing_2009,huang_reprogramming_2009}. The model is tractable for computation of full, discrete stochastic dynamics to within a small approximation error using matrix methods. Thus, the model provides a numerical benchmark for assessing the accuracy of the simulation method, before extension to larger systems where solution of the Chemical Master Equation (CME) is intractable. For the chosen parameters, the ExMISA model shows four peaks in the steady-state probability distribution (projected onto protein copy numbers, $n_a$ and $n_b$). Peaks in probability correspond to basins in the so-called quasipotential landscape, defined by $U=-\mathrm{ln}(\mathbf{\pi}(\mathbf{x}))$ (Fig~\ref{fig:MISAEx}). The four peaks/basins corresponds to four possible combinations of binarized $A/B$ gene expression: hi/hi, hi/lo, lo/hi, and lo/lo. These four phenotype-states arise due to the combination of balanced repression and self-activation in the network, and the slow kinetic parameters (Supplementary \nameref{S:ExMISAparam}) for transcription factor binding and unbinding to promoters that effect changes in individual gene-activity states between low and high expression rates~\cite{wang_epigenetic_2014,chu_markov_2017}. 

The WE-based simulation method enabled estimation of global dynamics of the ExMISA model. By redistributing computational resources from relatively high-probability to low-probability regions (see Methods), the WE method enabled uniform sampling of the quasipotential landscape, i.e., mapping basins (high-probability regions) along with high barriers (low probability regions) (Fig~\ref{fig:MISAEx}a). The simulation estimated individual steady-state bin-probabilities as low as $1.3\times 10^{-6}$ and showed good global agreement with the numerical CME benchmark (see Fig~\ref{fig:MISAEx} and Supplement, \nameref{S:flux_ExMISA_convg}). 

In addition to sampling global dynamics, the WE method can be used to estimate rate constants for individual, rare transitions of interest. The Mean First Passage Time of the global network switch from the center of one polarized phenotype-state to another, i.e., MFPT$_{X \rightarrow Y}$from protein a/b expression level hi/lo to lo/hi was estimated from WE to be $1.82\times 10^5$ (see Table~\nameref{S_MFPT_ExMISA}), in agreement with the CME result. 

\subsection*{Phenotype Transitions can be Approximated by Markovian Jumps, Enabling Construction of Coarse-Grained Models}

A network transition-matrix $\widetilde{\mathbf{T}}(\tau)$ over sampled bins ($N_\mathrm{bins}=300$) was constructed from WE sampling for ExMISA and used for subsequent analysis of global system dynamics. By comparison, a full network transition-matrix $\mathbf{T}(\tau)$ over the enumerated system state-space was constructed from the CME ($N=15876$, see Methods). The full, computed ($\mathbf{T}(\tau)$) and simulated ($\widetilde{\mathbf{T}}(\tau)$) transition-matrices showed qualitatively similar eigenvalue spectra with four dominant eigenvalues, indicating the presence of metastability (separation-of-timescales between intra-basin and inter-basin transitions) (Fig~\ref{fig:MISAEx}b). The slow system-timescales predicted by the full CME model corresponding to eigenvalues $\lambda_2,\lambda_3,\lambda_4$ were $t_2,t_3,t_4=6.8\times 10^4, \;4.2\times 10^4,\; 1.0\times 10^4$ respectively, in units of $k^{-1}$ where $k$ is the protein degradation rate (the Perron eigenvalue $\lambda_1=1$ is associated with the infinite-time (stationary) distribution). The corresponding values given by the WE-simulated  $\widetilde{\mathbf{T}}(\tau)$ were $6.1\times 10^4,\; 3.5\times 10^4, \;9.4\times 10^3$, respectively. These numbers demonstrate how the sampled $\widetilde{\mathbf{T}}(\tau)$ enables global approximation of slow system timescales to $<20\%$ relative error. Quantitative error in these values depends on both ``spectral'' (lagtime) and discretization error~\cite{prinz_markov_2011} (see \nameref{S:Voronoi_convergence}). In contrast, WE sampling in ``rate mode" (see Methods) enabled highly accurate estimation of MFPT$_{X \rightarrow Y}$ to within {\color{black} 2\% error} (\nameref{S_MFPT_ExMISA}).

According to the Markov State Model framework, the presence of timescale separation indicates that a simplified model, retaining a few coarse-grained metastable states with Markovian transitions among them, can reasonably approximate the full system dynamics. Using this approach, we label the metastable sets as \emph{phenotypes} accessible to the network, reasoning that a useful classification of cell phenotypes should be one that gives relatively stable, rather than transient, cell types. We apply the Markov State Model coarse-graining procedure to both the full $\mathbf{T}(\tau)$ and simulated $\widetilde{\mathbf{T}}(\tau)$, yielding similar results. The coarse sets (or metastable phenotype-states) in the reduced models for both cases are generated automatically, and map directly onto the four basins seen in the quasipotential landscape (i.e., the gene $A/B$ expression hi/hi, hi/lo, lo/hi, and lo/lo cell phenotypes). The reduced models are visualized by network graphs, in which node sizes are proportional to steady-state probability, and the thicknesses and lengths of edges are proportional to the transition probability between them (on lagtime $\tau$) (Fig~\ref{fig:MISAEx}c). Numerical values for the reduced models can be found in \nameref{S_MSMs}. The network graph can be considered to be an alternative representation of the global epigenetic landscape, which contains both stationary and dynamic information. (In contrast, the epigenetic landscape plotted as a quasipotential function does not explicitly contain dynamic information, due to non-gradient dynamics~\cite{wang_quantifying_2011}).

Validation of the coarse-grained model can be carried out according to the Chapman-Kolmogorov test~\cite{prinz_markov_2011}, which tests how well the relaxation dynamics initialized in the metastable phenotypes approximate the dynamics that are predicted either by the full model (CME) or simulated trajectories. According to this test, relaxation dynamics out of metastable phenotypes from WE sampling was predicted with error values between 0.02 and 0.12 for all phenotypes (\nameref{S:CK}). Together, these results indicate (i) that a Markovian model of phenotype transitions is a good approximation of the full system dynamics for the ExMISA model, and (ii) that the WE-simulation based computational pipeline predicts a quantitatively similar coarse-grained phenotype-network to the full CME model.

\subsection*{The Method Maps the Epigenetic Landscape and Identifies Dominant Phenotypes in a Pluripotency Network Model}

\begin{figure}[H]\includegraphics[width=\textwidth]{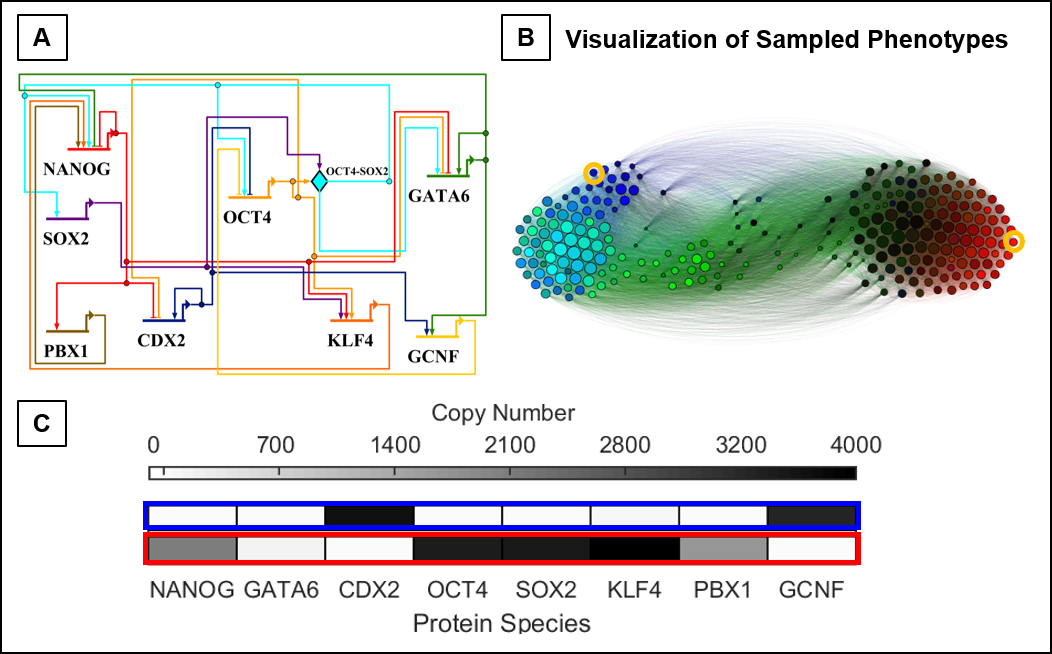}
\caption{{\bf Pluripotency Network Model and Simulation Results} A)Wiring diagram for the eight-gene pluripotency network model, adapted from~\cite{zhang_stem_2014}. Arrowheads represent positive interactions, while flat lines denote repression. B) Simulation results: state-transition graph of sampled network states. Circles represent aggregate gene-expression states sampled during the Weighted Ensemble simulation. Circle areas are proportional to the steady-state probability $\pi_i$ in each state according to $\mathrm{ln}(\gamma \pi_i)$ with scaling factor $\gamma = 3.4$. States are colored according to the gene expression levels of three of the genes; red, green, and blue correspond to high NANOG, GATA6, and CDX2 expression respectively, while black corresponds to low or no gene expression. Edges connecting the states indicate possible state-transitions, colored according to the originating state. The graph is produced using Gephi \cite{bastian_gephi:_2009} using a force-directed layout algorithm (Force Atlas), therefore short inter-state distances reflect higher probability of transitioning. C) Full protein compositions of two representative states, with either high CDX2 expression (blue) or high NANOG expression (red). States in (C) correspond to yellow circles in (B).
}
\label{Fig:NANOGintro}
\end{figure}

We apply the computational pipeline to a pluripotent fate-decision network from mouse Embryonic Stem Cells (mESCs) introduced by Zhang \emph{et al.} \cite{zhang_stem_2014} (Fig \ref{Fig:NANOGintro}A). The network comprises eight interacting genes: NANOG, GATA6, CDX2, SOX2, OCT4, GCNF, and PBX1. Three of these genes, NANOG, SOX2, and OCT4 have been suggested to maintain pluripotency\cite{chambers_nanog_2007}, and NANOG inhibits the expression of differentiation markers \cite{silva_nanog_2009}. The GATA6 and CDX2 genes have been used in experiments as markers of differentiation, with the GATA6 transcription factor being a marker of the primitive endoderm cell lineage, and the CDX2 transcription factor being a marker of the trophectoderm lineage \cite{hay_oct-4_2004}. 

Using the WE-based computational pipeline, we estimate $\widetilde{\mathbf{T}}(\tau)$ with a resolution of $N_\mathrm{bins}=250$. To visualize the global landscape as a graph network at this resolution, we plot the converged $\widetilde{\mathbf{T}}(\tau)$ using a force-directed automated graph layout \cite{bastian_gephi:_2009} (Fig \ref{Fig:NANOGintro}B). The barbell shape of the network reflects the broad antagonism between pluripotency and differentiation genes, which is a general feature of the overall network topology. At the same time, each ``pole'' comprises multiple distinct patterns of gene expression (seen in the graph as different colors with full compositions in Fig \ref{Fig:NANOGintro}C), hinting at the existence of multiple phenotypes associated with both pluripotency and lineage-specification. Moreover, the network representation reveals numerous links between pluripotent and differentiated states, pointing to both direct and indirect transitions, through a network of relatively transient intermediate states. 

\begin{figure}[H]\includegraphics[width=\textwidth]{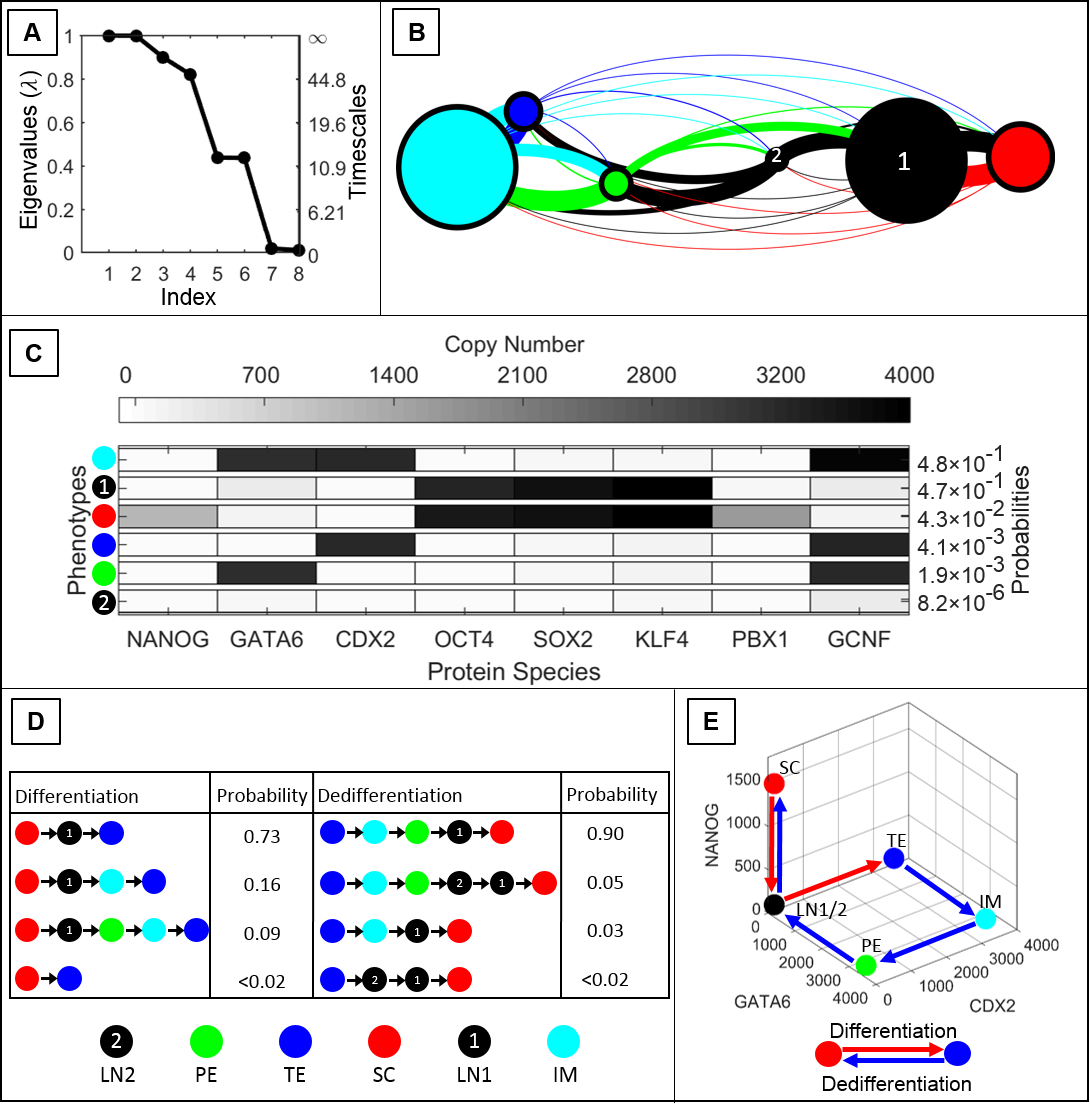}
\caption{{\bf Simulation Results for the Pluripotency Network (Parameter Set I). The Computational Pipeline Uncovers Six Metastable Phenotypes and Irreversible Phenotype Transitions.} A) Computed eigenvalue spectrum and global timescales indicating the presence of metastability in the network. The gap in the eigenvalue spectrum after the sixth eigenvalue suggests that a partitioning can be found into six metastable phenotypes. B) The coarse-grained network showing six algorithmically-identified phenotypes designated as Low NANOG 1 (LN1), Low NANOG 2 (LN2), Stem Cell (SC), Primitive Endoderm (PE), Trophectoderm (TE), and the Intermediate Cell (IM) state. C) The averaged gene expression levels (copy numbers) of each transcription factor for each phenotype and their respective steady-state probabilities. D) The four most probable transition pathways from the SC state to the TE state (differentiation) and from the TE state to the SC state (dedifferentiation). E) The highest probability transition paths projected onto three protein coordinates, NANOG, GATA6, and CDX2. Differentiation from SC to TE is visibly irreversible. }
\label{Fig:NANOGComp}
\end{figure}

To further analyze the global dynamics of the pluripotency network, we apply the Markov State Model coarse-graining framework. The simulated  $\widetilde{\mathbf{T}}(\tau)$ shows gaps in the eigenvalue spectrum after four and after six eigenvalues (Fig~\ref{Fig:NANOGComp}a). The corresponding approximate timescales are given by $t_2,\;t_3,\;t_4,\;t_5,\;t_6=1.1\times 10^{5},\;95,\;51,\;12,\;12$ ($k^{-1}$), respectively. These values, though only approximate, indicate the presence of a single long timescale process ($t_2$) corresponding to transfer between differentiated and pluripotent states, while transitions within those basins ($t_3$, etc.) occur at least four orders of magnitude more quickly. Applying the coarse-graining algorithm to achieve six clusters results in a reduced model (Fig~\ref{Fig:NANOGComp}b), with the clusters representing metastable phenotypes. The phenotypes can largely be distinguished in the subspace of NANOG, GATA6, and CDX2 expression levels; the differentiated phenotypes show expression of either GATA6 (primitive endoderm, PE), CDX2 (trophectoderm, TE), or both (denoted an intermediate cell type, IM). Phenotypes associated with pluripotency do not express high levels of GATA6 or CDX2, and may express high levels of NANOG (stem cell, SC). The coarse-grained model reveals two separate pluripotent phenotypes that are low in NANOG expression: one which expresses other pluripotent factors OCT4, SOX2, and KLF4 (``Low NANOG 1'' LN1), and one which has low expression of all factors (``Low NANOG 2'' LN2) (Fig~\ref{Fig:NANOGComp}c). Overall, these phenotypes broadly match experimentally-determined categories, coincide with steady-states of the stochastic model computed previously by a CME-approximation method\cite{zhang_stem_2014}, and coincide with phenotype-states identified in related pluripotency GRN models\cite{li_quantifying_2013-1}. The steady-state probabilities associated with the phenotypes are highly nonuniform, with $~95\%$ of the population divided nearly evenly between the IM and LN1 phenotypes, which are associated with differentiation and pluripotency, respectively. The LN2 state is rarest, comprising only $~8\times 10^{-4}\%$ of the population. Together, these results indicate that the clustering method identifies both common and exceedingly rare phenotypes in the \emph{in silico} cell population modeled by simulation trajectories. Furthermore, the automated method identifies both expected and novel phenotypes.

\subsection*{The Method Reveals Multiple, Irreversible Pathways for Phenotype Transitions in the Pluripotency Network}

Previously, Markov State Models constructed on the basis of Molecular Dynamics simulations were used to analyze the ensemble of distinct pathways of protein-folding~\cite{noe_constructing_2009}. Here, we utilize the coarse-grained model of phenotype transitions in the pluripotency GRN in a similar manner, to analyze pathways of cell differentiation and dedifferentiation. Using Transition Path Theory, the method identifies the pathways that carry the greatest fraction of net probability flux, among sequences associated with successful SC$\rightarrow$TE transitions (and reverse) (Fig~\ref{Fig:NANOGComp}d,e). Transition paths between the stem cell (SC) and PE phenotypes can be found in \nameref{S:f10_SC_PE}. For Parameter Set I, the method identifies three pathways encompassing $>98\%$ of the probability flux for both forward and reverse transitions. While the SC$\rightarrow$ TE transition is most likely to occur directly through the LN1 state (i.e., NANOG expression will shut off, followed by turning on CDX2), the reverse transition shows a different route through the IM and PE states (i.e., GATA6 expression turns on, then CDX2 turns off, then GATA6 turns off, and finally NANOG turns on). 

Dynamic analysis of the coarse-grained model, including analysis of transition paths, relies on the Markovian approximation for inter-phenotype transitions. In the pluripotency network, stochastic transitions between pluripotency (SC, LN1, LN2) and differentiation (TE, IM, PE) basins are infrequent relative to transitions within those basins, justifying the Markovian assumption, since the system equilibrates within those basins much more rapidly than inter-basin transitions occur. However, the Markovian assumption may be less accurate for describing intra-basin transitions between phenotypes, which occur much more frequently. Despite the coarse-grained model encompassing transitions on highly disparate timescales, the qualitative results of transition path analysis were validated by collected conventional simulation trajectories (not subject to any Markovian assumption), which identified the same dominant transition paths (\nameref{S:f10_path_vlid}). Overall, these results indicate that a stochastic excursion of a cell from the SC to TE phenotypes and back maps a cycle in gene-expression space, echoing previous studies indicating nonequilibrium dynamics in GRNs~\cite{wang_quantifying_2011,feng_2012}. The results further indicate that the Markov State Model, while a highly coarse-grained approximation, can provide an accurate estimation of inter-phenotype transition dynamics.

\subsection*{Cell Phenotype Landscape and Transition Dynamics are Sensitive to Kinetic Parameters}

We applied the computational pipeline to the pluripotency network using two different rate parameters sets (see \nameref{S1_File}), which differ in rates of transcription factor binding and unbinding to DNA. In line with previous studies\cite{feng_2012,tse_dna-binding_2015,chu_markov_2017}, we found that increasing the so-called adiabaticity (i.e., increasing $h$ and $f$, or the rates of TF-binding relative to protein production and degradation, Parameter Set II) led generally to rarer inter-phenotype transitions (see Table \ref{Tab:NANOGMFPT}). For example, in Parameter Set I, the Mean First Passage Time (MFPT) for transitions from SC $\rightarrow$ TE was calculated to be $1.36\times 10^{5}$ in units of $k^{-1}$, as compared to $8.13\times 10^{8}$ for Parameter Set II. The MFPTs of the reverse transition TE $\rightarrow$ SC for each set were $2.70\times 10^5$ and $5.82\times 10^9$, respectively (see Table \ref{Tab:NANOGMFPT} and \nameref{S_MFPT_SC_f10}). These differences in magnitude broadly reflect that moving toward the adiabatic regime leads to increased epigenetic barriers between phenotypes.

\begin{table}[H]
\centering
\begin{tabular}{|l+l | l | l| l|}
\hline
{\bf \emph{Transition}} & {\bf SC $\rightarrow$ LN(1)} & {\bf{LN(1)} $\rightarrow$ SC} & \textbf{SC $\rightarrow$ TE} & \textbf{TE $\rightarrow$ SC}\\
\thickhline
${Parameter \;Set\; I}\; (f=10)$  & $1.71 \times 10^1$ & $1.94 \times 10^2$ & $1.36\times 10^5$ & $2.70\times 10^5$\\
\hline
${Parameter\; Set \;II}\; (f=50)$ & $7.71\times 10^4$ & $1.28\times 10^4$ &   $8.13\times 10^8$&  $5.82\times 10^9$\\
\hline
\end{tabular}
\caption{{\bf Computed Mean First Passage Times (MFPTs) of Phenotype Transitions in the Pluripotency Network.} MFPTs are shown for transitions between the pluripotency (high NANOG) state (SC) and low NANOG expression states (LN(1)) (left columns) and for transitioning between the pluripotency state (SC) and the trophectoderm state (TE) (right columns), in units of the inverse transcription factor decay rate, $k^{-1}$. Transitions for Parameter Set I were computed using the WE method in rate mode while transitions for Parameter Set II were estimated from the sampled transition matrix. The definitions of SC and LN(1) are analogous to the high NANOG production (N$^{hi}$) and low NANOG production ($N^{lo}$) transitions measured in experiments~\cite{filipczyk_network_2015,hormoz_inferring_2016}. Increasing the adiabaticity (i.e., the rates of DNA-(un)binding, $h,f$), leads to rarer inter-phenotype transitions. The simulations also show that, within the same gene network for a given parameter set, inter-phenotype transition times span four orders of magnitude.} 
\label{Tab:NANOGMFPT}
\end{table}

\begin{figure}[H]\includegraphics[trim={0 2cm 0 0},clip, width = \textwidth]{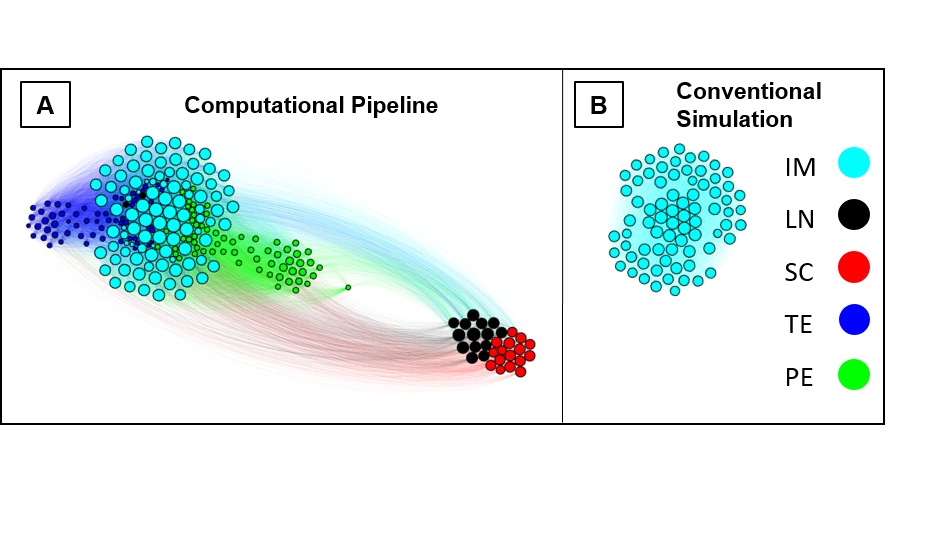}
\caption{{\bf The Rare-Event Sampling Pipeline Makes Rare States and Transitions Accessible to Simulation.} A) The global state-transition graph computed with the computational pipeline for the Pluripotency Network with rare transitions (Parameter Set II). The states are colored according to the coarse-grained (algorithmically-identified) phenotypes. In this parameter regime ($f=50$) the differentiated (TE, PE, IM) and pluripotent phenotypes are cleanly separated, reflecting exceedingly rare transitions between the two phenotypes (O($10^9$), see Table \ref{Tab:NANOGMFPT}). (B) States visited in conventional SSA simulation (using the same initialization, definitions, and placement as in (A)). In the conventional simulation, a transition out of the IM phenotype was never observed.}

\label{Fig:f50}
\end{figure}

\begin{figure}[H]\includegraphics[width=\textwidth]{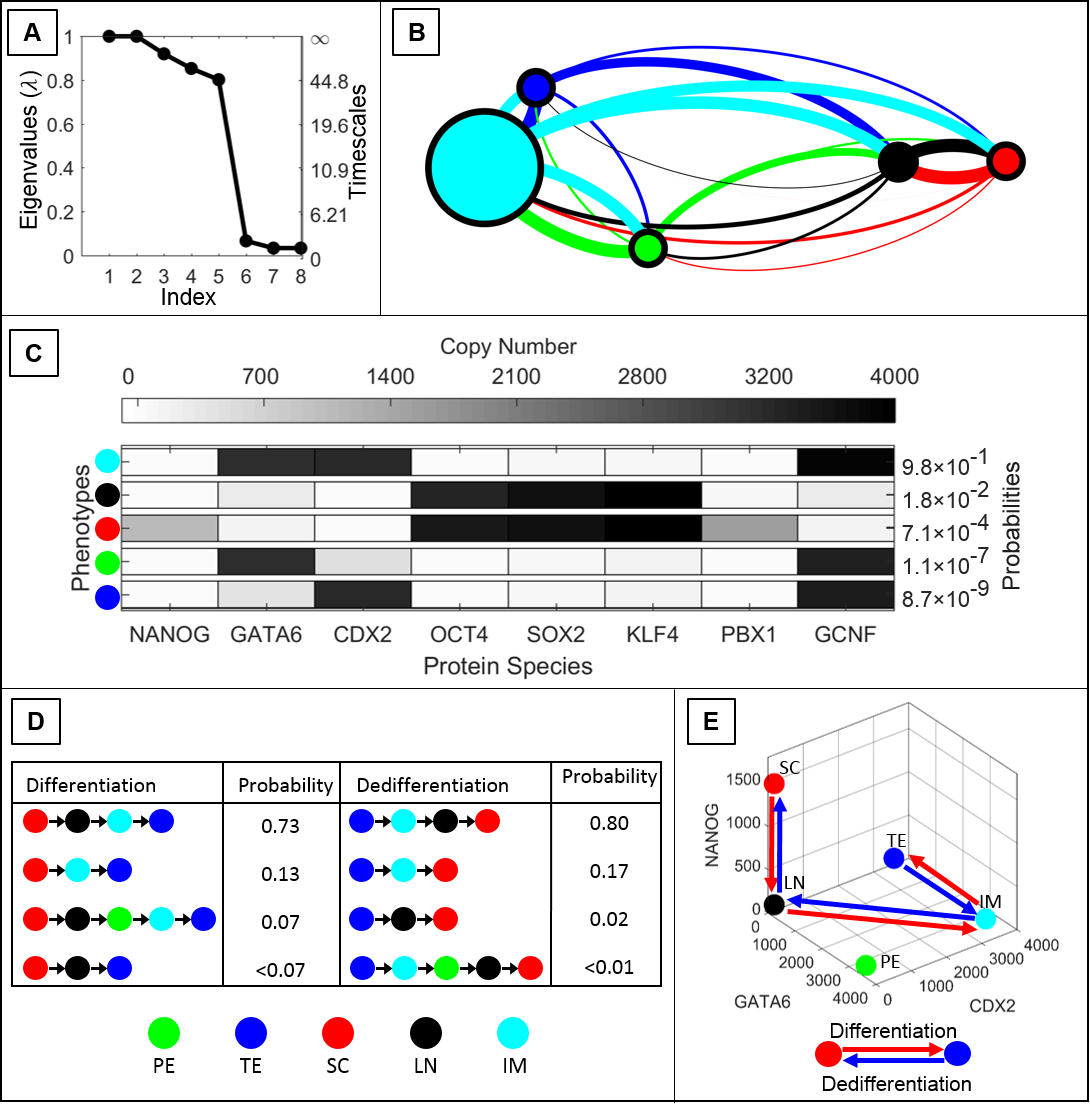}
\caption{{\bf Simulation Results for the Pluripotency Network (Parameter Set II). Changing DNA-Binding Kinetics Alters the Epigenetic Landscape.} A) Computed eigenvalue spectrum and global timescales. B) The coarse-grained Markov State Model showing five phenotypes corresponding to the LN1, SC, PPE, TE, and IM phenotypes of Parameter Set I. The majority of the steady state probability is in the IM phenotype $(0.98)$. C) The gene expression levels for each phenotype and their respective steady-state probabilities. D) The four most probable differentiation pathways between SC and TE phenotypes. E)The dominant pathways of (de)differentiation projected onto the GATA6, CDX2, and NANOG coordinates. The change in DNA-binding kinetics shows different transition dynamics from Parameter Set I. Here, the forward and reverse paths are the same.
}
\label{Fig:f50_2}
\end{figure}

In addition to generally slowing transitions, the increased adiabaticity of Parameter Set II gives rise to an epigenetic landscape structure that is distinct from that of Parameter Set I, with altered steady-state phenotype probabilities (Fig ~\ref{Fig:f50}a). The eigenvalue spectrum shows qualitatively distinct features as well, with a gap after five values (Fig~\ref{Fig:f50_2}a). As such, the Markov State Model framework identifies five dominant phenotypes in the network, which correspond broadly to those of Parameter Set I, except that only a single Low-NANOG (LN) phenotype is identified (Fig~\ref{Fig:f50_2}b). Most of the steady-state probability is contained in the IM state (Fig~\ref{Fig:f50_2}c). In addition to altering the transition rates and relative phenotype probabilities, the kinetic parameters altered the dynamics of differentiation and dedifferentiation. The two likeliest pathways of forward (and reverse) SC $\rightarrow$ TE transitions follow the same route through LN and IM phenotypes (Fig~\ref{Fig:f50_2}d,e). Alternative differentiation pathways of forwards (and reverse) SC $\rightarrow$ PE transitions can be found in ~\nameref{S:f50_SC_PE}. These results indicate that, while the same GRN model with different kinetic parameters may give rise to qualitatively similar phenotypes, they differ in quantitative stationary and dynamic features, including relative steady-state probabilities, transition times, and likeliest transition pathways.

\subsection*{Efficiency of Rare-Event Sampling Compared to Conventional SSA}

\begin{figure}[H]\includegraphics[width=\textwidth]{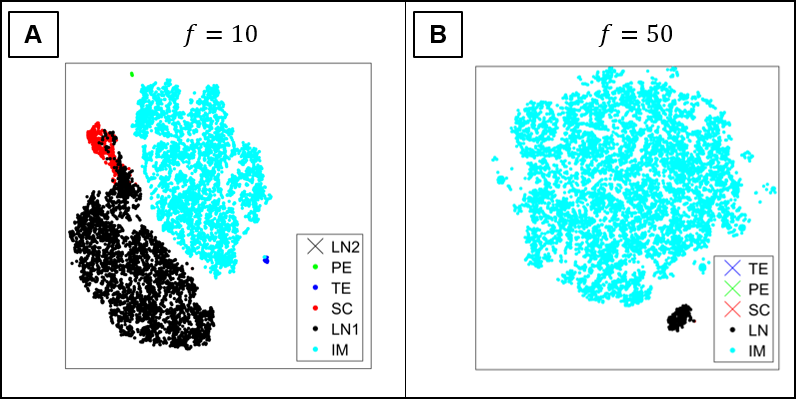}
\caption{{\bf ``Synthetic" Cell Population Data Computed by the Rare-Event Sampling Pipeline, Visualized with a Single Cell Visualization Method (tSNE). } A) tSNE visualization of $15000$ simulated 'cells' (replicas) drawn from WE sampling for Parameter Set I. Each cell is colored according to its phenotype after the coarse-graining. The population is heavily dominated by the IM and LN phenotypes, though all other phenotypes are sampled with the exception of the rare LN2 phenotype. b) tSNE visualization of Parameter Set II simulation data. Only the LN and IM phenotype-states are sampled in a synthetic population of size 15000.
}
\label{fig:tSNE}
\end{figure}

Phenotype transitions that are relatively rare can be difficult to observe with conventional SSA simulation. We compared simulated landscapes (based on estimated $\widetilde{\mathbf{T}}(\tau)$) from the computational pipeline to those obtained from an equivalent (large) number of SSA simulation steps (Fig~\ref{Fig:f50}a,b). Additional comparisons of synthetic cell populations using tSNE visualization reflect the rarity of phenotypes and phenotype transitions (Fig ~\ref{fig:tSNE}, \nameref{S:tSNE_alt}). This comparison revealed that the WE-based method uncovers multiple phenotypes and associated transitions that are invisible to conventional simulation due to the rarity of exiting metastable basins. Quantitative estimates of efficiency gains for WE have often been based on comparing the number of simulation steps required to estimate a desired quantity (such as a rate constant) using WE versus conventional simulation~\cite{donovan_efficient_2013}. Treating $\widetilde{\mathbf{T}}(\tau)$ as the desired output (as it contains holistic dynamic information for the system), we estimate the efficiency gain of our pipeline by computing:
\begin{eqnarray}
\label{eq:efficiency}
E=\frac{\textrm{Sim. steps to estimate }\widetilde{\mathbf{T}}(\tau) \textrm{, Conv.} }{\textrm{Sim. steps to estimate } \widetilde{\mathbf{T}}(\tau) \textrm{, WE}}.
\end{eqnarray}
However, it is often difficult to acquire the required number of steps for conventional simulation, so an approximate lower bound for the denominator can be estimated according to:
\begin{eqnarray}
\label{eq:sumMFPTs}
\left[ \textrm{Sim. steps to estimate }\widetilde{\mathbf{T}}(\tau) \textrm{, Conv.}\right] \gtrapprox\sum_{i,j}\mathrm{MFPT}_{i,j},
\end{eqnarray}
where simulation steps and transition times are measured using the same time-unit (here, $k^{-1}$). That is, the denominator is the sum over the MFPTs of transitions between each pair of states (bins), $i,j$, where $i,j=1...N_{\mathrm{bins}}$. This approximation is based on the rationale that one requires simulation time $\mathcal{O}$(MFPT) to observe at least one transition between a given pair of states. From the WE-estimated transition-matrix $\widetilde{\mathbf{T}}(\tau)$, estimates of the MFPT for transitions between any pair of states (bins) can be obtained using Eq ~\ref{eq:MFPT}. According to Eq ~\ref{eq:efficiency}, we estimate that our pipeline provided efficiency gains of 3000 for ExMISA (Fig.~\ref{fig:MISAEx}), 200 for Pluripotency Parameter Set I (Fig.~\ref{Fig:NANOGintro}), and $4\times 10^7$ for Parameter Set II (Fig.~\ref{Fig:f50_2}). These numbers show that the pipeline affords a significant speedup over conventional simulation in providing global dynamic information. The numbers further show that the efficiency gain is most pronounced for the Pluripotency network with exceedingly rare inter-phenotype transitions.

\section*{Discussion}
In this work, we present a method for efficient, automated computation of epigenetic landscapes, metastable phenotypes, and phenotype-transition dynamics of stochastic GRN models. Our computational pipeline was inspired by studies of metastability and barrier-crossing in Molecular Dynamics, and our application of the pipeline to cell-scale networks addresses a number of current challenges  for stochastic GRN dynamics. First, it overcomes the curse-of-dimensionality of complex models, by leveraging available rule-based modeling tools for stochastic biochemical networks~\cite{faeder_rule-based_2009}. Second, it overcomes the challenge of efficiently simulating stochastic systems with rare events, by using enhanced Weighted Ensemble rare-event sampling~\cite{huber_weighted-ensemble_1996}. Third, it addresses the challenge of extracting and interpreting essential dynamics of complex systems on the basis of simulated trajectories, by using the Markov State Model framework~\cite{noe_constructing_2009} to automatically generate a compact, approximate representation of global system dynamics. Combining these tools into a unified pipeline provides an automated means of computing and visualizing essential stationary and dynamic properties of stochastic GRNs, including the number and identities (i.e. state-space mapping) of metastable phenotypes, their steady-state probabilities, and most-likely pathways of inter-phenotype transitions and their transition rates. By advancing the capability to compute and interpret hypothesized or experimentally-derived stochastic GRN models, the method can yield insight into how ``local'' stochastic, molecular processes involved in epigenetic regulation affect ``global'' dynamics such as phenotypic stability and fate-transitions in cells. Moreover, it can help close the gap between dynamic, molecular-detailed models of gene regulation and cell-population level experimental data, to inform rational cell reprogramming strategies. 

\subsection*{Insights from the Pluripotency Network Simulations}
We used the pluripotency network as a model system to develop and demonstrate the simulation approach, but the results also yielded biological insights. For example, the simulations revealed a hierarchical structure of the epigenetic landscape. The network--exhibiting 5-6 metastable phenotypes--occupies a limited subspace from the vast possible gene combinations (e.g., $2^8=256$ possible distinct on/off combinations of gene expression states). The dominant feature of the global landscape is a high barrier/slow timescale between pluripotent and differentiated phenotypes. Within each of these categories, further sub-states were identified. The model revealed multi-timescale dynamics of phenotype transitions; the pluripotency network showed relatively rapid transitions between phenotype-states that differed in the expression-level (high vs. low) of a single gene, e.g. the high NANOG to low NANOG transition, whereas phenotype transitions involving a change in expression level of seven genes, e.g. the SC macrostate to the TE macrostate, occurred five orders of magnitude more slowly on average. 

While the accessible phenotypes appear broadly similar across parameter sets, the relative stability and transition dynamics among phenotypes were sensitive to kinetic parameters governing transcription factor binding/unbinding. A global change in these parameters (affecting all individual transcription factor-DNA interactions equally) changed the shape of the landscape, altering the relative steady-state probabilities of different phenotypes and the likely transition pathways linking them. The DNA binding parameters capture the local epigenetic mechanisms that enable/disable transcription factors from accessing regulatory elements. A global rate change nevertheless has a varying influence on different genes because the number of regulators differs, as does the molecular logic by which activators and repressors exert combinatorial control on different genes. These results echo findings that global modification of chromatin regulators often have lineage-specific effects \cite{Constantinides}. These results highlight both the need for, and the challenge, of informing cell reprogramming strategies with quantitative network models, as they suggest that the dynamic response of cellular networks to perturbations is governed by the detailed kinetics of molecular regulatory mechanisms, which are generally difficult to parameterize. 

\subsection*{Dynamic Definition of Cell Phenotype}
The Markov State Model framework implicitly imposes a dynamic definition of cell phenotypes; the number of phenotypes was determined using spectral gap-analysis, and the coarse-graining algorithm automatically identified metastable aggregates (i.e., grouped sampled network states into larger clusters). This is different from the classifications of phenotypes that are generally used in analyzing experimental data, where gene expression or marker levels are often used to categorize cells. However, experiments have also revealed the potential need for a dynamic definition of cell phenotype, based not only on single-timepoint measurements of gene expression or phenotype-markers, but also on information from past or future timepoints\cite{kalmar_regulated_2009,filipczyk_network_2015}. For example, Filipczyk \emph{et al.}~\cite{filipczyk_network_2015} identified distinct subpopulations within a compartment of NANOG-negative cells in mESCS, which differed in their propensity to re-express NANOG. At the same time, fluctuations between low- and high-NANOG expressing cells were not necessarily associated with any functional state change. The Markov State Model approach, based on kinetic/dynamic coarse-graining, thus provides a quantitative approach for classifying phenotype-states that is both completely generalizable rather than \emph{ad hoc} (it requires no \emph{a priori} knowledge or designation of markers/genes) and is in line with these recent experiments revealing the need for a dynamic definition of phenotype. 

\subsection*{Timescales of Stochastic Phenotype Transitions}
Markovian transitions (i.e., memoryless ``hops'') among cell phenotypes have been observed experimentally: examples include transitions among phenotypes in cancer cells, as measured by flow cytometry~\cite{gupta_stochastic_2011}, and among pluripotency-states in mESCs, as measured by time-lapse microscopy of fluctuating gene expression~\cite{singer_dynamic_2014,filipczyk_network_2015,hormoz_inferring_2016}. The compact nature of these data-inferred networks--showing hops among a limited set of broad phenotypes--suggests that the computed MSM framework advanced in this study provides an appropriate level of resolution at which to analyze GRN dynamics and may serve as a useful tool for comparing models to experimental data. 

Experimental studies have quantified the timescales of Markovian transitions between NANOG-high and NANOG-low states in mESCs~\cite{filipczyk_network_2015,hormoz_inferring_2016}. From Hormoz \emph{et al.}, the probability of transitioning from NANOG-high to NANOG-low in mESCs is $0.02$ per cell cycle, while that of the reverse transition is $0.08$. These values represent a relatively rapid transition rate, since NANOG expression is known to be particularly dynamic~\cite{silva_nanog_2009}. Similarly, plasticity has been observed in cancer cells where quantitative estimates of stochastic cell transitions between a stem cell cancer cell phenotype to a basal cancer cell phenotype were observed to be roughly on the order of 0.01 to 0.1 per cell cycle~\cite{gupta_stochastic_2011}. We can translate our model results to approximate biological timescales: the degradation rate, which sets the timeunit for model results (i.e., $k$ is taken to be 1) was experimentally determined to be on the order of a few hours (in the E14 mouse embryonic stem cell line, the half-lives of NANOG, OCT4, and SOX2 are approximately 4.7, $>$ 6, and 1.6 hours, respectively \cite{abranches_generation_2013}). Assuming that degradation is unimolecular, $k = \ln(2)/t_{[NANOG]1/2}$, and the half-life of NANOG, $t_{[NANOG]1/2} = 5$ hours, the degradation rate is $k = 0.1$. Using a mESC cell cycle time of 12 hours \cite{wakayama_mice_1999}, the simulations for Parameter Set I then predict NANOG-high to NANOG-low transitions occurring with a rate of 0.03 
per cell cycle, and of $3\times 10^{-3}$ 
for the reverse. For Parameter Set II, the computed rates were $8\times 10^{-6}$ 
and $5\times 10^{-5}$, respectively. Comparison of these computed and experimental rates of NANOG transitions indicates that Parameter Set I ($f=10$) is more in line with experimental observations, while Parameter Set II ($f=50$) gives transition rates that are three orders of magnitude too slow. These results are in agreement with previous findings from theoretical studies that GRNs in pluripotency networks operate in a so-called ``weakly-adiabatic'' regime \cite{sasai_time_2013,zhang_stem_2014,tse_dna-binding_2015}, in which the timescale of DNA-binding by transcription factors is on the order of transcription factor production and degradation.

\subsection*{Comparison to Other Models and Computational Approaches}
A number of theoretical studies have elucidated dynamics of stochastic molecular-detailed GRN models (i.e., models that include molecular fluctuations and regulatory mechanisms, in contrast to Boolean models\cite{chang_systematic_2011}). These studies have largely focused on small 1- or 2-gene motifs[\cite{aurell_epigenetics_2002,kepler_stochasticity_2001,sasai_stochastic_2003,feng_2012,ma_small-number_2012,tse_dna-binding_2015,ge_stochastic_2015}], but recent years have seen extension of stochastic methods to studies of more complex, experimentally derived GRN models encompassing $\mathcal{O}(10)$ genes. For example, determination of global dynamic properties of such networks has been achieved by combining information from long stochastic simulations of discrete models~\cite{sasai_time_2013,li_quantifying_2013-1}, or of continuum SDE models, in combination with path integral approaches~\cite{wang_epigenetic_2014,li_quantifying_2015}. The pluripotency network studied herein was developed by Zhang and Wolynes~\cite{zhang_stem_2014}; in their work, the authors developed a continuum approximation to the Chemical Master Equation that enabled quantitative construction of the epigenetic landscape. Here, we present an alternative approach that is unique in two major aspects: (1) the use of stochastic simulations (i.e., SSA~\cite{gillespie_exact_1977}), which is enabled by use of the WE rare-event sampling algorithm, and (2) the automated Markov State Model framework for designating phenotypes and constructing a coarse-grained view of the epigenetic landscape. While we utilize a different framework (that of coarse-grained, discrete stochastic models) from Zhang and Wolynes to approximate and interpret dynamics, our results are broadly consistent with theirs. For example, the dominant identified phenotypes we found are the same as in their work (the only exception being the exceedingly rare LN2 phenotype identified by the coarse-graining algorithm for Parameter Set I).

\subsection*{Current Challenges and Future Directions}
Our approach is uniquely suited to extracting global dynamics information for stochastic systems with metastability, using simulations. An advantage of this approach is that both the WE and coarse-graining algorithms are``dynamics-agnostic"~\cite{donovan_efficient_2013}, meaning that they can be applied to any type of stochastic dynamics framework. In the context of computational biology, our pipeline could be extended to other types of stochastic biochemical systems, such as systems with hybrid discrete-continuum dynamics~\cite{hepp_adaptive_2015}, systems with spatial heterogeneity~\cite{donovan_unbiased_2016}, or multi-level models \cite{maus_rule-based_2011}. In addition to this flexibility, simulation-based methods have the advantage of being able to leverage existing, widely-used open-source packages, which in turn facilitate model specification and model sharing. For example, BioNetGen~\cite{faeder_rule-based_2009} can interpret models specified in the Systems Biology Markup Language.

Several challenges and potential weaknesses with the pipeline exist, both with regard to sampling rare events, and in determining an appropriate coarse-grained model. Potential challenges with the WE algorithm have been described elsewhere\cite{donovan_unbiased_2016,zuckerman_weighted_2017}, and include the difficulty of determining a binning that captures slow degrees of freedom and the existence of time-correlations between sampled iterations of the simulation, which can impede unbiased sampling. The Voronoi-based binning procedure we employ here is related to a number of similar approaches~\cite{dickson_nonequilibrium_2009, zhang_weighted_2010, dickson_wexplore:_2014,tse_dna-binding_2015}, and has the advantage of effectively tiling a high-dimensional space without the need for \emph{a priori} knowledge. However, in practice, according to others and our own studies, the method is effective up to about 10 degrees of freedom. These challenges are the subject of continued study.

\newpage


\section*{Supporting information}

 For more information, see \nameref{S1_File} and \nameref{S2_File}.

\paragraph{File S1}
\label{S1_File}
{\bf Description of network models, kinetic parameters, and weighted ensemble parameters}  

\paragraph{File S2}
\label{S2_File}
{\bf Pseudo-code for the computational pipeline.}

\paragraph{Table S1}
\label{S:ExMISAparam}
{\bf ExMISA Network Parameters}

\paragraph{Table S2}
\label{S:pluripotency_network}
{\bf Pluripotency Network}

\paragraph{Table S3}
\label{S:pluripotency_network_param}
{\bf Pluripotency Network Parameters}

\paragraph{Table S4}
\label{S:WE_param}
{\bf Weighted Ensemble Simulation Parameters}

\paragraph{Table S5}
\label{S_MSMs}
{\bf Transition Matrices of Metastable Phenotype Clusters (MSMs)}

\paragraph{Table S6}
\label{S_MFPT_ExMISA}
{\bf Computed Mean First Passage Times in the ExMISA Network--Comparison of Different Methods} 

\paragraph{Table S7}
\label{S_MFPT_SC_f10}
{\bf Computed Mean First Passage Times of Inter-Phenotype Transitions in the Pluripotency Network (Parameter Set I)}

\paragraph*{Fig S1}
\label{S:Voronoi_movement}
{\bf Movement of Voronoi Centers during weighted ensemble sampling} 

\paragraph{Fig S2}
\label{s:FSP_error}
{\bf Error in computed steady-state probability as a function of $N$, the number of protein states retained in the state-space truncation.}

\paragraph{Fig S3}
\label{S:flux_ExMISA_convg}
{\bf Convergence of the flux of the transition between the polarized phenotype-states in the ExMISA network}

\paragraph{Fig S4}
\label{S:Voronoi_convergence}
{\bf Convergence of the slowest implied timescale  $t_2$ with increasing number of sampling regions (bins) and increasing lagtime $\tau$}

\paragraph{Fig S5}
\label{S:CK}
{\bf The Chapman-Kolmogorov test on the four Markov State Model phenotypes of the sampled ExMISA network.}

\paragraph*{Fig S6}
\label{S:f10_SC_PE}
{\bf{Pathway decomposition for the SC $\rightarrow$ PE transition for $f=10$}}

\paragraph*{Fig S7}
\label{S:f10_path_vlid}
{\bf Validation of the SC $\rightarrow$ TE transition pathway calculated through weighted ensemble simulation}

\paragraph*{Fig S8}
\label{S:f10_alt}
{\bf Reproducibility of the weighted ensemble sampling of the pluripotency network.}

\paragraph*{Fig S9}
\label{S:f50_SC_PE}
{\bf{Pathway decomposition for the SC $\rightarrow$ PE transition for $f=50$}}

\paragraph*{Fig S10}
\label{S:tSNE_alt}
{\bf tSNE plot of replica populations with rescaled weight ($\ln(P_s) + 30$)}

\paragraph{Fig S11}
\label{S:flux_NANOG_convg}
{\bf Convergence of the flux of the TE $\rightarrow$ SC transition in the pluripotency network with $f=10$}

\paragraph{Fig S12}
\label{S:CG_partitioning}
{\bf Difference in Coarse-Grained clustering for the 2-gene ExMISA cell decision network studied through the numerical benchmark (top) and the WE sampling pipeline (bottom).}

\section*{Acknowledgments}
We thank the administrators of the University of California, Irvine High-Performance Computing Cluster and we thank Jun Allard for helpful discussions.

\newpage


%
%
%
\bibliography{AdaptiveWE_MSMpaper}

\raggedright
\setlength{\parindent}{0.5cm}
\textwidth 5.25in 
\textheight 8.75in



\title{\Huge Supporting Information}
\date{}
\maketitle


\section{ExMISA Network}
Two-gene network with Mutual Inhibition, Self-Activation, and exclusive transcription factor binding.
\begin{flushleft}
\ce{A_{00} + 2a <=>[h_a][f_a] A_{10}}\\
\ce{A_{00} + 2b <=>[h_r][f_r] A_{01}}\\
\ce{B_{00} + 2b <=>[h_a][f_a] B_{10}}\\
\ce{B_{00} + 2a <=>[h_r][f_r] B_{01}}\\
\ce{A_{00} ->[g_0] A_{00} + a}\\
\ce{B_{00} ->[g_0] B_{00} + b}\\
\ce{A_{01} ->[g_0] A_{01} + a}\\
\ce{B_{01} ->[g_0] B_{01} + b}\\
\ce{A_{10} ->[g_1] A_{10} + a}\\
\ce{B_{10} ->[g_1] B_{10} + b}\\
\ce{a ->[k] 0}\\
\ce{b ->[k] 0}
\end{flushleft}

\section{Pluripotency network}
There are eight genes (encoding transcription factors) in the pluripotency network. Transcription factors bind as homodimers with the exception of the OCT4-SOX2 heterodimer. Only three transcription factors interact with their own gene, CDX2, NANOG, and GATA6. Transcription factors bind as dimers with the rate $h$ and unbind with the rate $f$. When a gene is bound by any activator and no repressors, it expresses at a rate $g_{on}$, otherwise, it expresses at a rate $g_{off}$. The only exception is NANOG, which must be bound by all three of its activators and no repressors to be activated.

\raggedright
\setlength{\parindent}{0.5cm}
\textwidth 5.25in 
\textheight 8.75in




\title{\Huge{Pseudocode for the Computational Pipeline}}

\section{Weighted Ensemble Exploration Mode}
\begin{enumerate}
\item Format the reaction network into a BioNetGen file. 
\item Choose $M_\mathrm{targ}$, the target number of replicas per sampling region, and $N_\mathrm{bins}$, the target number of sampling regions or bins.
\item Initialize $M_\mathrm{targ}$ replicas in a single starting location.
\item Simulate replicas for a simulation time $\tau_\mathrm{WE}$. Replicas are simulated in parallel.
\item Chose $N_\mathrm{bins}$ new bin positions.
\begin{enumerate}
\item Chose one random replica as the first new bin position.
\item Chose the replica furthest from the set of bin positions to be the next new bin position.
\item Repeat (b) until $N_\mathrm{bins}$ new positions have been chosen.
\end{enumerate}
\item Perform the WE step.
\begin{enumerate}
\item For a given bin, if the number of replicas in the bin is less than $M_\mathrm{targ}$, split the replica with the largest weight into $n$ equally weighted replicas until there are $M_\mathrm{targ}$ replicas. 
\item For a given bin, if the number of replicas in the bin is greater than $M_\mathrm{targ}$, combine the weight of $n$ replicas and randomly chose one to receive the combined weight such that there are $M_\mathrm{targ}$ replicas in the bin.
\item Repeat (a) or (b) for each sampling bin
\end{enumerate}
\item Repeat steps 4-6 for a chosen number of simulation steps.
\end{enumerate}

\section{Transition-Matrix Mode}
\begin{enumerate}
\item Start from the end of exploration mode. 
\item Simulate replicas for a time $\tau_\mathrm{WE}$.
\item Collect weights transferred from bin $i$ to bin $j$ over the simulation period $\tau$ into a transition matrix.
\item Perform the WE step.
\item Repeat steps 2-4 for a chosen number of simulation steps.
\end{enumerate}

\section{Rate-Estimation Mode}
\begin{enumerate}
\item Start from the end of exploration mode. 
\item Label replicas as having most recently visited region of interest $X$ or visited region of interest $Y$.
\item Simulate replicas for a time $\tau_\mathrm{WE}$.
\item Collect weights transferred from $X$ to $Y$ over the simulation period $\tau$ and change the replica label as necessary.
\item (Optional) Chose $N_\mathrm{bins}$ new sampling regions.
\item Perform the WE step.
\item Repeat steps 3-6 for a chosen number of simulation steps.
\end{enumerate}
\section{Coarse-Graining Procedure}
\begin{enumerate}
\item Find the left-eigenvalues and eigenvectors of the row-stochastic transition matrix calculated from transition-matrix estimation mode.
\begin{enumerate}
\item The probability distribution of the system is estimated by the left-eigenvector associated with the eigenvalue $\lambda = 1$
\end{enumerate}
\item Perform the PCCA+ algorithm using MSMBuilder software to cluster the $N_\mathrm{bins}$ sampling regions into macrostates. MSMBuilder software outputs a Markov State Model of the reaction network.
\item Use transition path analysis (using PyEMMA software) on the resulting MSM to obtain parallel transition paths or estimate the rate of transitioning between any two states. 
\item Gephi 0.7 is used to visualize the row-stochastic transition matrix and the MSM.
\end{enumerate}

\raggedright
\setlength{\parindent}{0.5cm}
\textwidth 5.25in 
\textheight 8.75in




\setcounter{figure}{0}    
\section{Supporting Figures}

\begin{figure}[H]
\centering\includegraphics[width=1.0 \linewidth]{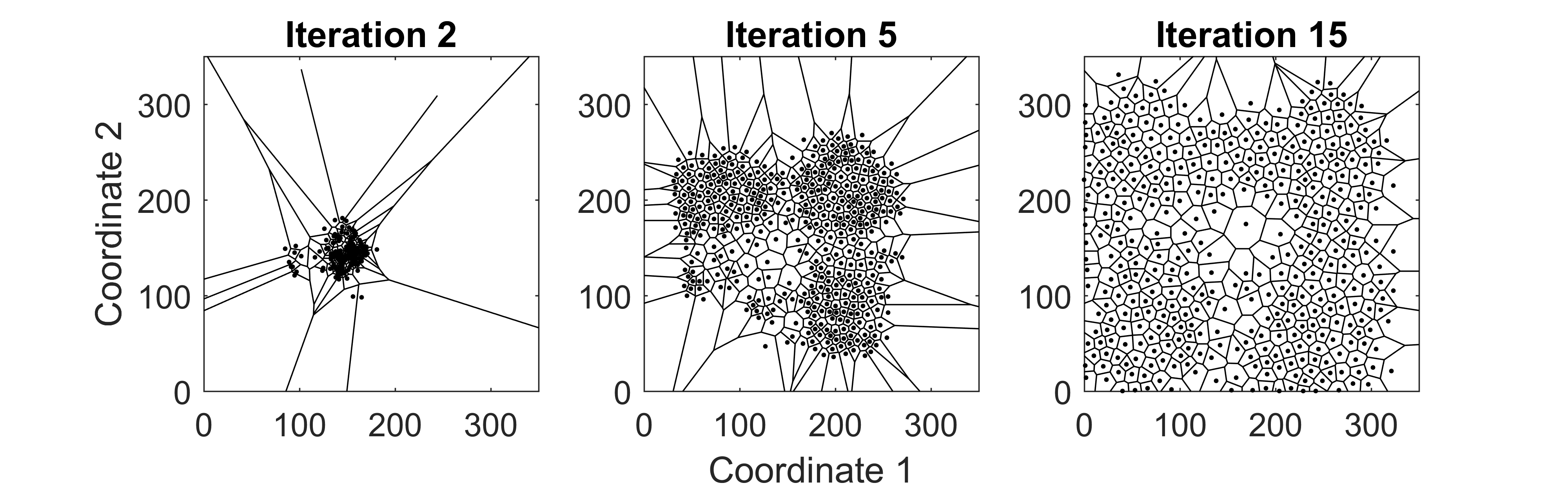}
\caption{{\bf Movement of Voronoi Centers during weighted ensemble sampling.} Starting from the left are shown three successive iterations of the adaptive WE simulation for a representative network.}
\label{S:Voronoi_movement}
\end{figure}

\begin{figure}[H]
\centering\includegraphics[width=1.0\linewidth]{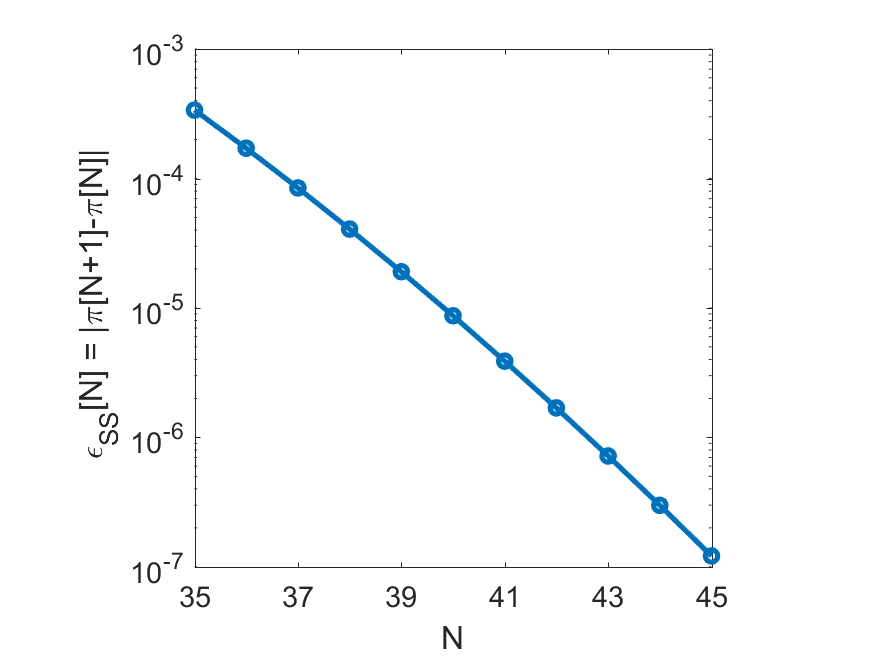}
\caption{{\bf Error in computed steady-state probability as a function of $N$, the number of protein states retained in the state-space truncation.} $N$ corresponds to the maximum allowed copy-number of transcription factors $a$ and $b$ in the ExMISA network. For a truncation to $N$, probability flux between states with $n_a$, $n_b \le N$ and states with $n_a$, $n_b>N$ is assumed to be 0 (i.e., the boundaries of the state-space are reflective). The error $\epsilon_SS[N]$ is defined by $\sum_i\left| \pi[N+1]-\pi[N] \right|$, where $i$ runs over all enumerated states of the state-space with truncation to $N+1$ (all states outside the boundary have probability 0). That is, the error is computed as the sum of the absolute difference between steady-state probabilities for each state, comparing $\pi[N]$ (steady-state probability computed with truncation to $N$) to $\pi[N+1]$ (truncated to $N+1$).}
\label{s:FSP_error}
\end{figure}

\begin{figure}[H]
\centering\includegraphics[width=1 \linewidth]{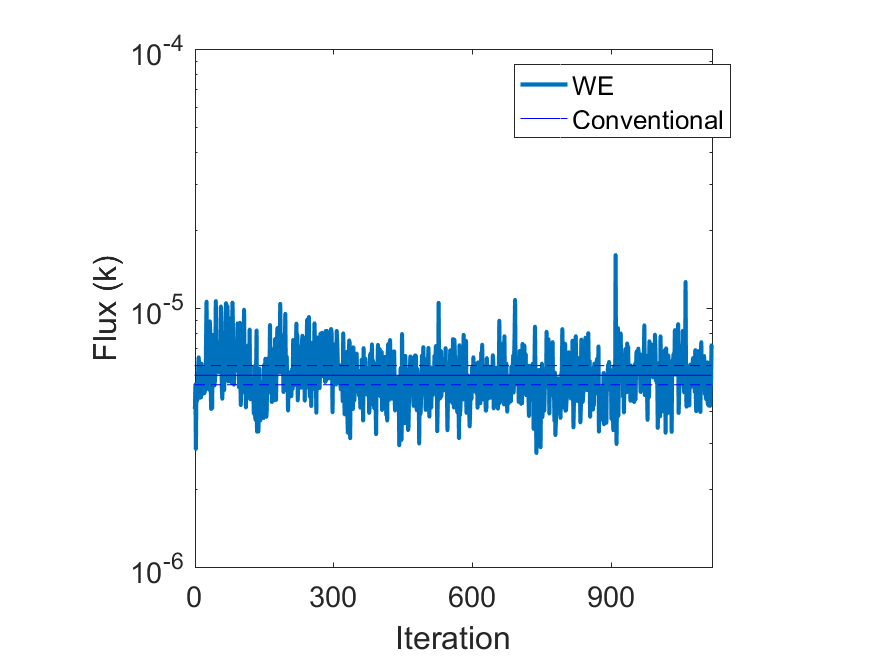}
\caption{{\bf Convergence of the flux of the transition between the polarized phenotype-states in the ExMISA network.} The 5\% and 95\% confidence intervals for the long conventional simulation are shown in dotted blue lines. The flux between the a/b hi/lo and lo/hi phenotypes was calculated using WE sampling with parameters: $\tau=200$, 300 bins, and 50 replicas per bin. The system was sampled for 1100 iterations of $\tau$.
}
\label{S:flux_ExMISA_convg}
\end{figure}

\begin{figure}[H]
\centering\includegraphics[width=0.6 \linewidth]{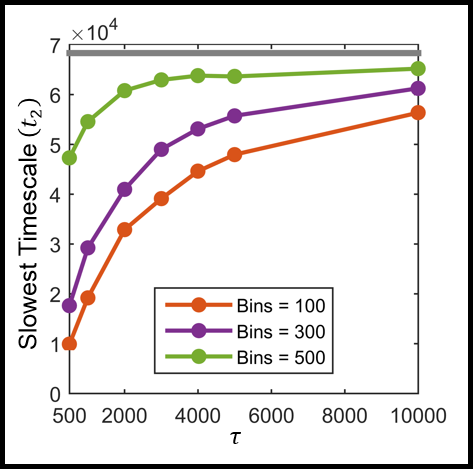}
\caption{{\bf Convergence of the slowest implied timescale  $t_2$ with increasing number of sampling regions (bins) and increasing lagtime $\tau$.} The lagtime calculated using the truncated CME is shown in gray. The accuracy of the WE approximation increases monotonically with increasing bin number and lagtime.
}
\label{S:Voronoi_convergence}
\end{figure}

\begin{figure}[H]
\centering\includegraphics[width=0.6 \linewidth]{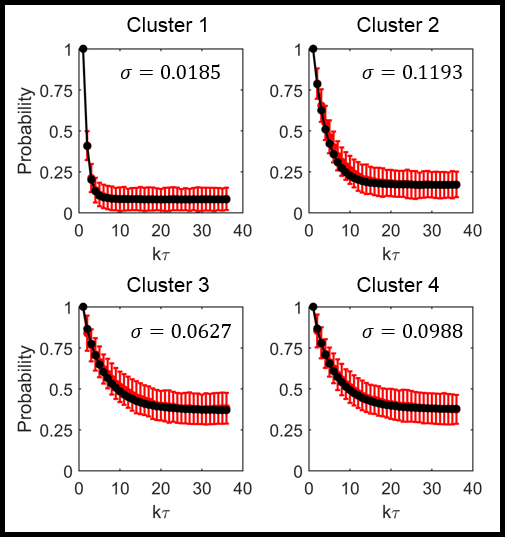}
\caption{ {\bf The Chapman-Kolmogorov test on the four Markov State Model phenotypes of the sampled ExMISA network.} The relaxation curves and variance of a 1000$\tau$ trajectory are shown in red. The relaxation curve predictions from the MSM transition matrix is shown in black. The total error $\sigma$ is measured as the 2-norm containing the differences between the two estimates of the relaxation curve. }
\label{S:CK}
\end{figure}

\begin{figure}[H]
\centering\includegraphics[width=1 \linewidth]{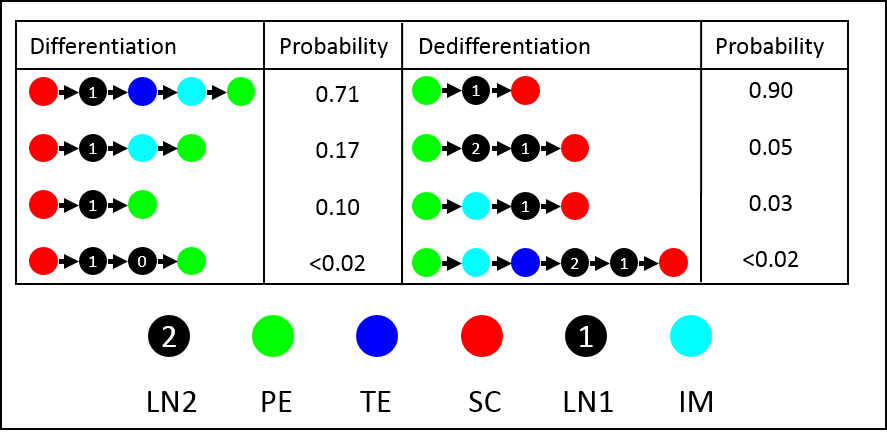}
\caption{\bf{Pathway decomposition for the SC $\rightarrow$ PE transition for $f=10$}}
\label{S:f10_SC_PE}
\end{figure}

\begin{figure}[H]
\centering\includegraphics[width=1 \linewidth]{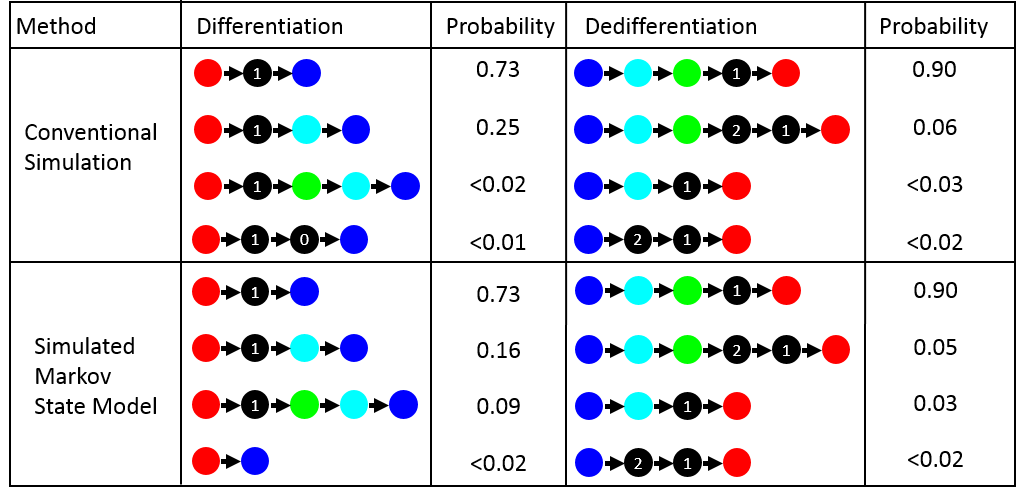}
\caption{{\bf Validation of the $\mathbf{SC} \rightarrow \mathbf{TE}$ transition pathway calculated through weighted ensemble sampling.} The parallel transition pathways are compared against those calculated from a single long conventional simulation.}
\label{S:f10_path_vlid}
\end{figure}

\begin{figure}[H]
\centering\includegraphics[width=1 \linewidth]{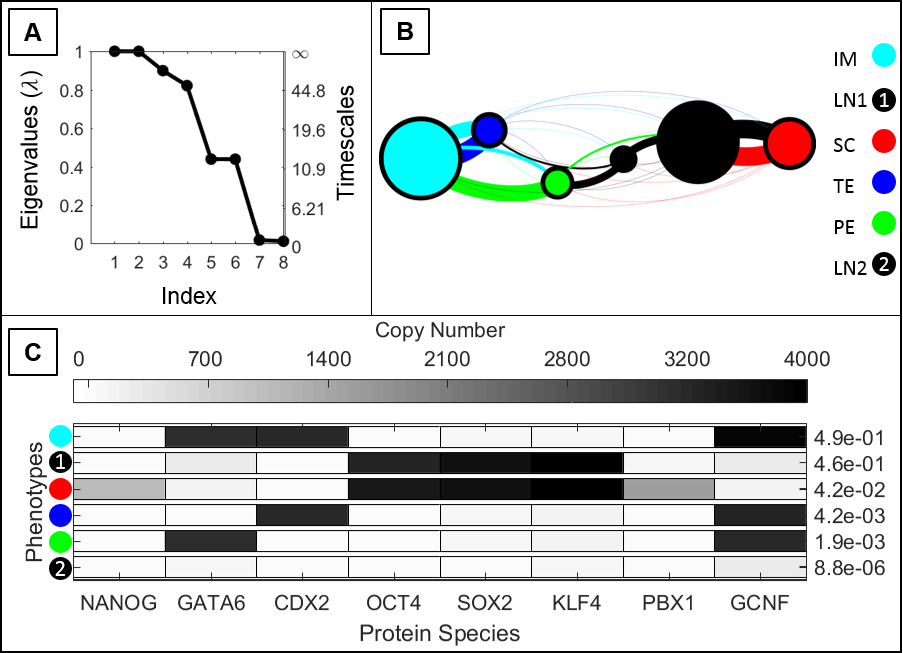}
\caption{{\bf Reproducibility of the weighted ensemble sampling of the pluripotency network.} The second WE sampling of $f=10$ parameter set was initialized in the same manner as the first. A) Eigenvalues and timescales. B) MSM C) Macrostate compositions}
\label{S:f10_alt}
\end{figure}

\begin{figure}[H]
\centering\includegraphics[width=1 \linewidth]{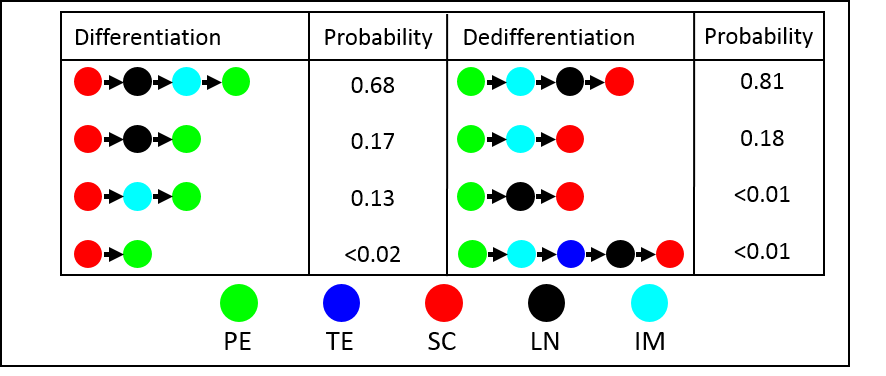}
\caption{\bf{Pathway decomposition for the SC $\rightarrow$ PE transition for $f=50$}}
\label{S:f50_SC_PE}
\end{figure}

\begin{figure}[H]
\centering\includegraphics[width=1\linewidth]{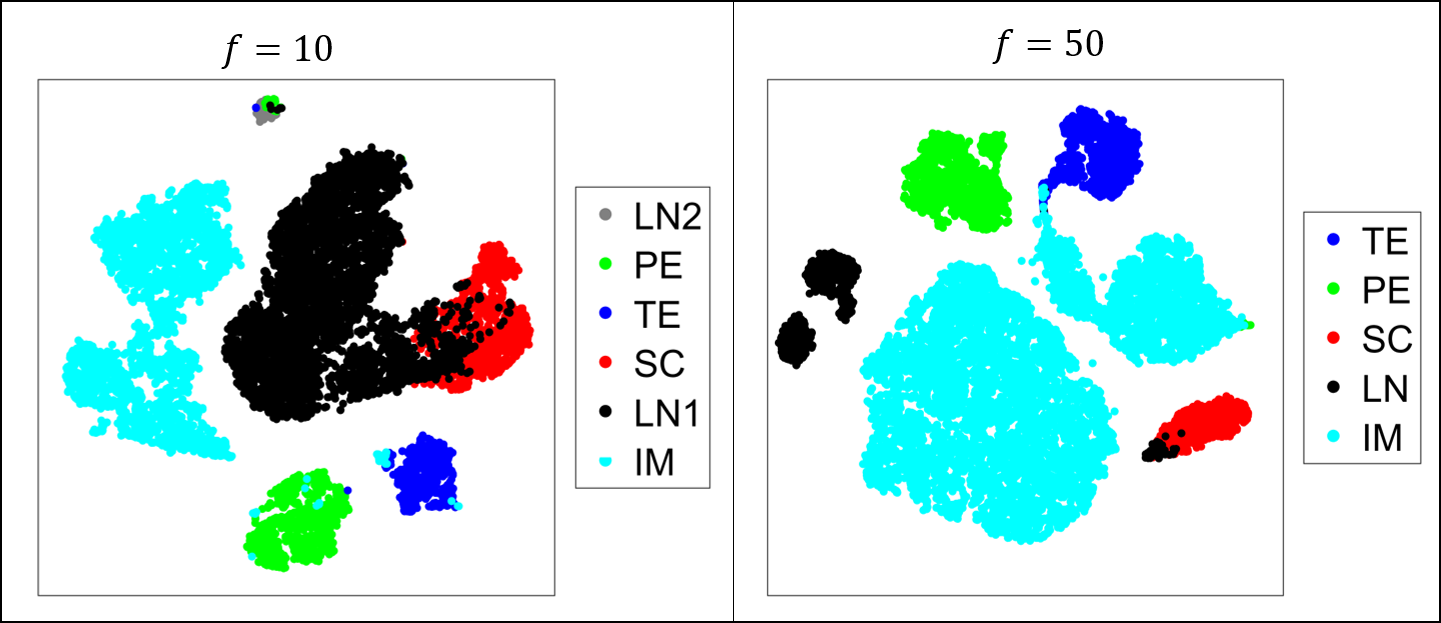}
\caption{{\bf tSNE plot of replica populations with rescaled weight ($\ln(P_s) + 30$)}}
\label{S:tSNE_alt}
\end{figure}

\begin{figure}[H]
\centering\includegraphics[width=1 \linewidth]
{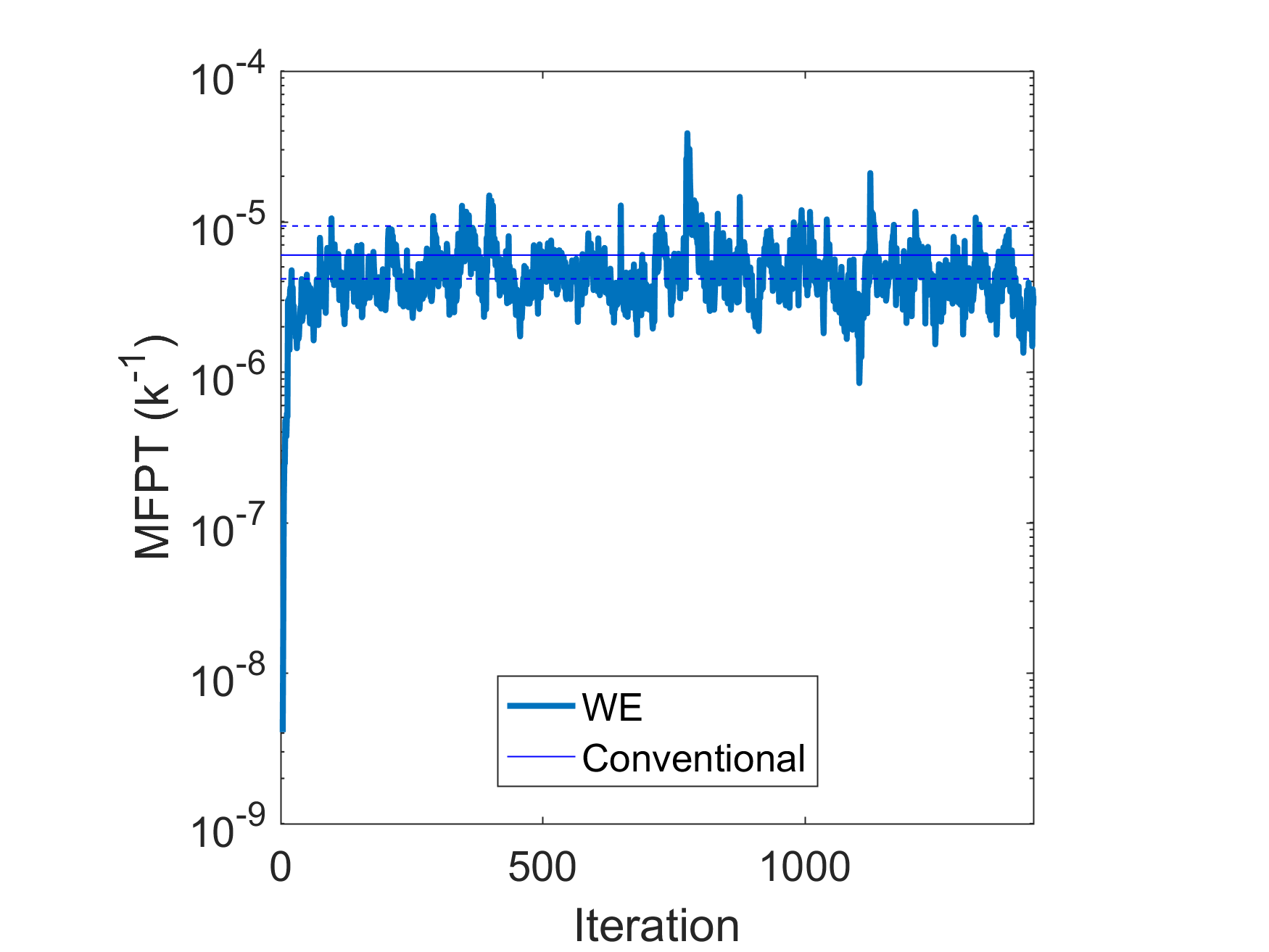}
\caption{{\bf Convergence of the flux of the TE $\rightarrow$ SC transition in the pluripotency network with $f=10$}. The 5\% and 95\% confidence intervals for the long conventional simulation are shown in dotted blue lines. The flux was calculated using WE sampling with parameters: $\tau=50$, 250 bins, and 100 replicas per bin. The system was sampled for 1400 iterations of $\tau$.}
\label{S:flux_NANOG_convg}
\end{figure}

\begin{figure}[H]
\centering\begin{adjustwidth}{-2.25in}{0in}\includegraphics[]{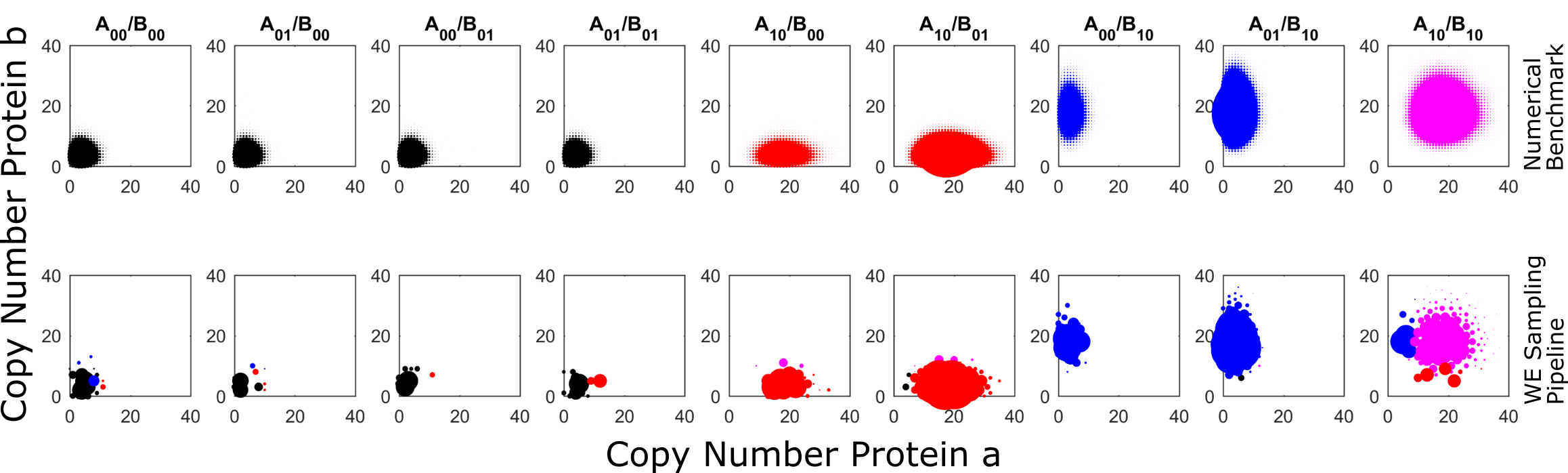}
\end{adjustwidth}
\caption{{\bf Difference in Coarse-Grained clustering for the 2-gene ExMISA cell decision network studied through the numerical benchmark (top) and the WE sampling pipeline (bottom).} The color of each coarse-grained phenotype cluster corresponds to the expression level of protein a/b: lo/lo (black), hi/lo (red), lo/hi (blue), hi/hi (magenta). All enumerated state phenotypes are sized proportionally to their probability for each of the nine gene configurations for the numerical benchmark, while only the centers of each sampling region are shown for the WE sampling computational pipeline. While the centers of the sampling regions are mostly well separated according to their gene configuration, the phenotype states assigned to the sampling region extend across multiple gene configurations due to the choice of euclidean distance metric in assigning phenotype states to sampling regions.}
\label{S:CG_partitioning}
\end{figure}

\raggedright
\setlength{\parindent}{0.5cm}
\textwidth 5.25in 
\textheight 8.75in



\section{Supporting Tables}
\setcounter{table}{0}

\begin{table}[H]
\begin{adjustwidth}{-2.25in}{0in} 
\centering
\caption{
{\bf ExMISA Network Parameters}}
\begin{tabular}{|l+l | l |}
\hline
{\bf ExMISA Parameters} & {\bf Value in $[1/k]$} & \bf Description\\
\thickhline
$g_0$ & 4.0 & basal/ repressed expression rate  \\
\hline
$g_1$ & 18.0 & activated expression rate  \\
\hline
$h_a$   & $1\times 10^{-5}$ & binding rate of activator \\
\hline
$h_r$   & $1\times 10^{-1}$ & binding rate of repressor  \\
\hline
$f_a$   & $1\times 10^{-5}$ & unbinding rate of activator \\
\hline
$f_r$   & $1$ & unbinding rate of repressor \\
\hline
k & 1 & transcription factor degradation rate  \\
\hline
\end{tabular}
\begin{flushleft} Parameters of the ExMISA network
\end{flushleft}
\label{ExMISAParam}
\end{adjustwidth}
\end{table}

\begin{table}[H]
\begin{adjustwidth}{-2.25in}{0in} 
\centering
\caption{
{\bf Pluripotency Network}}
\begin{tabular}{|l + l | l|}
\hline
 \textbf{Gene} & \textbf{Activators} & \textbf{Repressors}\\
\thickhline
$PBX1$ & NANOG & --- \\
\hline
$CDX2$ & CDX2 & NANOG, OCT4\\
\hline
$NANOG$ & PBX1, OCT4-SOX2,KLF4 & NANOG, GATA6\\
\hline
$GATA6$ & GATA6, OCT4-SOX2 & NANOG, OCT4\\
\hline
$GCNF$ & CDX2, GATA6 & --- \\
\hline
$KLF4$ & NANOG, OCT4, SOX2 & --- \\
\hline
$OCT4$ & OCT4-SOX2 & GCNF, CDX2\\
\hline
$SOX2$ & OCT4-SOX2 & --- \\
\hline

\end{tabular}
\begin{flushleft} Interaction rules for genes in the pluripotency network.
\end{flushleft}
\label{NANOGInteractions}
\end{adjustwidth}
\end{table}

\begin{table}[H]
\begin{adjustwidth}{-2.25in}{0in} 
\centering
\caption{
{\bf Pluripotency Network Parameters}}
\begin{tabular}{| l + l | l | l |}
\hline
{\bf Parameter} & Set I & Set II & \bf Description\\
\thickhline
$g_{off}$ & 100 & 100 & basal/ repressed expression rate  \\
\hline
$g_{on}$ & 3900 & 3900 & activated expression rate  \\
\hline
$h$   & $1\times 10^{-5}$ & $5\times 10^{-5}$& binding rate of transcription factor \\
\hline
$f$   & $10$ & $50$ & unbinding rate of transcription factor  \\
\hline
$k$ & 1 & 1 &transcription factor degradation rate  \\
\hline
\end{tabular}
\begin{flushleft} Parameters of the pluripotency network in units of $k^{-1}$.
\end{flushleft}
\label{NANOGParam}
\end{adjustwidth}
\end{table}

\begin{table}[H]
\begin{adjustwidth}{-2.25in}{0in} 
\centering
\caption{
{\bf Weighted Ensemble Simulation Parameters}}
\begin{tabular}{|l+l | l | l | l|}
\hline
{\bf WE Parameters} & {\bf ExMISA }  & {\bf ExMISA } & {\bf Pluripotency } & {\bf Pluripotency }\\
{} & {\bf (Voronoi} & \bf{(Transition } & \bf{$\mathbf{f=10}$ and $\mathbf{f=50}$} &  \bf{$\mathbf{f=10}$ and $\mathbf{f=50}$} \\
{} & {\bf Movement)}  & {\bf Matrix Mode)} & \bf{(Voronoi Movement)}& \bf{(Transition Matrix Mode)}\\
\thickhline
$\tau$ & $10000$ & $10000$ & $10$ & $10$\\
\hline
\emph{Simulation regions} & $300$& $300$ & $250$& $250$\\
\hline
\emph{Replicas per region} & $100$& $100$ & $500$ & $500$\\
\hline
\emph{Iterations} & $60$ & $600$ & $60$ & $600$\\
\hline
\end{tabular}
\begin{flushleft} WE parameters for all networks.  
\end{flushleft}
\label{WEParam}
\end{adjustwidth}
\end{table}

\newpage
\begin{table}[H]
\begin{adjustwidth}{-2.25in}{0in} 
\centering
\caption{
{\bf Transition Matrices of Metastable Phenotype Clusters (MSMs)}}
\begin{tabular}{|l+l | l | l | l |}
\hline
\multicolumn{5}{|c|}{\bf{ExMISA Network}}\\
\thickhline
 & {\bf State 1 (lo/lo)} & {\bf State 2 (hi/hi)} & {\bf State3 (lo/hi)} & {\bf State 4 (hi/lo)}  \\
\thickhline
\emph{State 1 (lo/lo)} & $8.96\times 10^{-1}$ & $5.63\times 10^{-4}$ & $5.29\times 10^{-2}$ & $5.05\times 10^{-2}$\\
\hline
\emph{State 2 (hi/hi)} & $1.24\times 10^{-4}$ & $9.53\times 10^{-1}$ & $2.22\times 10^{-2}$ & $2.45\times 10^{-2}$\\
\hline
\emph{State 3 (lo/hi)} & $1.16\times 10^{-2}$ & $1.99\times 10^{-2}$ & $9.68\times 10^{-1}$ & $5.52\times 10^{-4}$\\
\hline
\emph{State 4 (hi/lo)} & $1.05\times 10^{-2}$ & $2.15\times 10^{-2}$ & $6.78\times 10^{-4}$ & $9.67\times 10^{-1}$\\
\hline
\end{tabular}
\begin{tabular}{|l+l|l|l|l|l|l|}
\hline
\multicolumn{7}{|c|}{\bf{Pluripotency Network Parameter Set I}}\\
\thickhline
 & {\bf State 1 (LN2)} & {\bf State 2 (PE)} & {\bf State3 (TE)} & {\bf State 4 (SC)}& {\bf State 5 (LN1)} & {\bf State 6 (IM)} \\
\thickhline
\emph{State 1 (LN2)} & $2.97\times 10^{-1}$& $2.50\times 10^{-1}$& $6.04\times 10^{-2}$& $3.47\times 10^{-3}$& $2.68\times 10^{-1}$ & $2.07\times 10^{-2}$\\
\hline
\emph{State 2 (PE)} & $1.42\times 10^{-3}$& $8.90\times 10^{-1}$& $3.06\times 10^{-4}$& $1.83\times 10^{-4}$& $1.11\times 10^{-2}$ & $9.66\times 10^{-2}$\\
\hline
\emph{State 3 (TE)} & $1.91\times 10^{-4}$& $2.11\times 10^{-4}$& $8.03\times 10^{-1}$& $3.03\times 10^{-7}$& $1.00\times 10^{-4}$ & $1.96\times 10^{-1}$\\
\hline
\emph{State 4 (SC)} & $5.09\times 10^{-12}$& $4.30\times 10^{-6}$& $9.34\times 10^{-6}$& $4.30\times 10^{-1}$& $5.70\times 10^{-1}$ & $1.39\times 10^{-7}$\\
\hline
\emph{State 5 (LN1)} & $2.06\times 10^{-6}$& $5.16\times 10^{-6}$& $3.53\times 10^{-5}$& $5.16\times 10^{-2}$& $9.48\times 10^{-1}$ & $8.20\times 10^{-6}$\\
\hline
\emph{State 6 (IM)} & $2.72\times 10^{-7}$& $4.14\times 10^{-4}$& $1.64\times 10^{-3}$& $1.01\times 10^{-9}$& $1.36\times 10^{-6}$ & $9.98\times 10^{-1}$\\
\hline
\end{tabular}

\begin{tabular}{|l+l|l|l|l|l|}
\hline
\multicolumn{6}{|c|}{\bf{Pluripotency Network Parameter Set II}}\\
\thickhline
 & {\bf State 1 (TE)} & {\bf State 2 (PE)} & {\bf State3 (SC)} & {\bf State 4 (LN)}& {\bf State 5 (IM)}  \\
\thickhline
\emph{State 1 (TE)} & $8.05\times 10^{-1}$& $2.66\times 10^{-6}$& $1.31\times 10^{-6}$& $2.92\times 10^{-3}$& $1.92\times 10^{-1}$ \\
\hline
\emph{State 2 (PE)} & $3.09\times 10^{-7}$& $9.20\times 10^{-1}$& $3.13\times 10^{-7}$& $1.65\times 10^{-4}$& $7.98\times 10^{-2}$ \\
\hline
\emph{State 3 (SC)} & $4.70\times 10^{-8}$& $2.21\times 10^{-7}$& $8.23\times 10^{-1}$& $1.77\times 10^{-7}$& $2.38\times 10^{-6}$\\
\hline
\emph{State 4 (LN)} & $5.25\times 10^{-9}$& $4.92\times 10^{-8}$& $6.86\times 10^{-3}$& $9.93\times 10^{-1}$& $5.02\times 10^{-7}$\\
\hline
\emph{State 5 (IM)} & $1.60\times 10^{-9}$& $8.22\times 10^{-9}$& $2.22\times 10^{-9}$& $1.01\times 10^{-8}$& $9.99\times 10^{-1}$ \\
\hline
\end{tabular}

\begin{flushleft} Markov State Models of metastable phenotype-cluster transitions found through the computational pipeline for all three simulated networks. There are four different combinations of a/b protein expression levels in the coarse-grained phenotype network of the ExMISA network: lo/lo, hi/hi, lo/hi, and hi/lo. The steady-state probabilities of the lo/lo, hi/hi, lo/hi, and hi/lo cell phenotypes are predicted by the computational pipeline to be $1.71\times 10^{-2},\; 7.67\times 10^{-2},\;3.71\times 10^{-1}, \;3.80\times 10^{-1}$, respectively. The steady state probabilities of the six and five coarse-grained phenotype networks in the pluripotency network Parameter Set I and Parameter Set II can be found in figures 4 and 6, respectively.
\end{flushleft}
\label{S_MSMs}
\end{adjustwidth}
\end{table}

\begin{table}[H]
\begin{adjustwidth}{-2.25in}{0in} 
\centering
\caption{
{\bf Computed Mean First Passage Times in the ExMISA Network--Comparison of Different Methods}}
\begin{tabular}{|p{4cm}+p{4cm}|l|l|l|}
\hline
{\bf Method} & {\bf Start/End State} & {\bf MFPT } & \bf{5\% Confidence} &  \bf{95\% Confidence}\\
\thickhline
\emph{CME - numeric benchmark} & Basin centers & $1.84\times 10^5$  & --- & ---  \\
\hline
\emph{Conventional SSA simulation} & Basin centers & $1.82\times 10^5$ &$1.67\times 10^5$ & $1.98\times 10^5$\\
\hline
\emph{WE - rate mode} & Basin centers & $1.82\times 10^5 $ & $1.78\times 10^5$ & $1.85\times 10^5$  \\
\hline
\emph{WE - transition matrix mode} & Coarse-grained polarized phenotype & $2.34\times 10^5$  & --- & ---\\
\hline
\emph{Coarse-grained phenotype network} & Coarse-grained polarized phenotype & $1.70\times 10^5$  & --- & --- \\
\hline
\end{tabular}
\begin{flushleft} Computed Mean First Passage Times (MFPTs, time-units $k^{-1}$) of the ExMISA network, using different computation methods. For each row, $\mathrm{MFPT}_{XY}=\mathrm{MFPT}_{YX}$ due to symmetry in the network, and the start- and end-state ($X$ and $Y$) for the transition are defined either with respect to distance from the centers of the polarized phenotype basins, or in terms of aggregated states in the coarse-grained phenotype definition. For basin centers, State $X$ is defined as a hypersphere of radius 1 centered around the state vector [4,16,0,0,1,0,1,0], corresponding to the species: [$a$,$b$,$A_{00}$,$A_{10}$,$A_{01}$,$B_{00}$,$B_{10}$,$B_{01}$]. State Y is a hypersphere centered around [16,4,0,1,0,0,0,1]. For the coarse-grained phenotype definition, states correspond to the polarized a/b hi/lo and lo/hi phenotypes.

\end{flushleft}
\label{S_MFPT_ExMISA}
\end{adjustwidth}
\end{table}

\begin{table}[H]
\begin{adjustwidth}{-2.25in}{0in} 
\centering
\caption{
{\bf Computed Mean First Passage Times of Inter-Phenotype Transitions in the Pluripotency Network (Parameter Set I)}}
\begin{tabular}{| l + l | l | l | l |}
\hline
{\bf Method } & \multicolumn{4}{ |c| }{\bf Transition}\\ 
\hline 
& {\bf SC $\rightarrow$ LN1} & {\bf LN1 $\rightarrow$ SC} & {\bf SC $\rightarrow$ TE} & {\bf TE $\rightarrow$ SC}\\

\thickhline
$CME$ & ---  & --- & --- & --- \\
\hline
$Conventional SSA$ & $1.85(1.75,2.33)\times 10^1$ & $2.69(2.30,3.43)\times 10^2$ & $1.65(1.07,2.34)\times 10^5 $ & $3.59(2.50,5.18)\times 10^5$ \\
\hline
\emph{Weighted Ensemble}  & $1.71(1.64,1.78)\times 10^1 $ & $1.94(1.85,2.05)\times 10^2 $ & $1.36(1.02,1.77)\times 10^5$ & $2.70(2.48,2.91)\times 10^5 $  \\
\emph{Rate Mode} &&&&\\
\hline
\emph{Weighted Ensemble} & $2.09 \times 10^1$ & $4.13\times 10^2$ & $2.19\times 10^5$ & $2.21\times 10^5$\\
\emph{Transition Matrix Mode} &&&&\\
\hline
$MSM$ & $2.23\times 10^1$ & $3.89\times 10^2$ & $2.06\times 10^5$ &  $2.20\times 10^5$\\
\hline
\end{tabular}
\begin{flushleft} The Mean First Passage Times of NANOG Fluctuation and (De)differentiation in the pluripotency network calculated using $\tau=10$ and $f=10$. The MFPT reported for the WE is a block average over the last 500 iterations. The MFPT and standard deviation are found for transitioning between the stem cell phenotype (SC) and the pluripotent phenotype with low NANOG expression (LN1) (analogous to high NANOG production (N$^{hi}$) and low NANOG production ($N^{lo}$) transitions measured in experiments) and for transitioning between the stem cell phenotype (SC) and the trophectoderm phenotype (TE), calculated on the timescale of the protein degradation rate $k$. The SC, TE, and LN1 regions of interest (ROI) are defined as the SC, TE, and LN1 phenotypes derived from the MSM reduction of the sampled transition matrix.
\end{flushleft}
\label{S_MFPT_SC_f10}
\end{adjustwidth}
\end{table}

\end{document}